\pdfoutput=1

\documentclass[aoas]{imsart}\usepackage[]{graphicx}\usepackage[]{color}
\makeatletter
\def\maxwidth{ %
  \ifdim\Gin@nat@width>\linewidth
    \linewidth
  \else
    \Gin@nat@width
  \fi
}
\makeatother

\usepackage{url}
\usepackage{cleveref}

\definecolor{fgcolor}{rgb}{0.345, 0.345, 0.345}

\usepackage{framed}
\makeatletter
 {\par\unskip\endMakeFramed%
 \at@end@of@kframe}
\makeatother

\definecolor{shadecolor}{rgb}{.97, .97, .97}
\definecolor{messagecolor}{rgb}{0, 0, 0}
\definecolor{warningcolor}{rgb}{1, 0, 1}
\definecolor{errorcolor}{rgb}{1, 0, 0}
\newenvironment{knitrout}{}{} 

\usepackage{alltt}

\usepackage{dsfont}
\usepackage{amsthm,amsmath,amssymb,natbib}
\usepackage{xspace,soul}
\usepackage{graphicx}

\usepackage[margin=1.45in]{geometry}

\startlocaldefs

\newcommand{\Ex}{\mathbb{E}}

\newcommand{\R}{\textsf{R}\xspace}
\newcommand{\pkg}[1]{\texttt{#1}\xspace}

\definecolor{orange}{rgb}{1, 0.5, 0}

\def\balpha{\pmb{\alpha}}

\endlocaldefs
\IfFileExists{upquote.sty}{\usepackage{upquote}}{}
\begin{document}

\begin{frontmatter}

\title{How often does the best team win?\\A unified approach
to understanding randomness in North American sport}
\runtitle{Randomness in sport}

\author{\fnms{Michael J.} \snm{Lopez}\corref{}\ead[label=e1]{mlopez1@skidmore.edu}}
\address{\printead{e1}}
\affiliation{Skidmore College}

\and
\author{\fnms{Gregory J.} \snm{Matthews}\ead[label=e2]{gmatthews1@luc.edu}}
\address{\printead{e2}}
\affiliation{Loyola University Chicago}

\and
\author{\fnms{Benjamin S.} \snm{Baumer}\ead[label=e3]{bbaumer@smith.edu}}
\address{\printead{e3}}
\affiliation{Smith College}

\runauthor{Lopez, Matthews, Baumer}

\begin{abstract}

Statistical applications in sports have long centered on how to best separate signal (e.g. team talent) from random noise. However, most of this work has concentrated on a single sport, and the development of meaningful cross-sport comparisons has been impeded by the difficulty of translating luck from one sport to another. In this manuscript, we develop Bayesian state-space models using betting market data that can be uniformly applied across sporting organizations to better understand the role of randomness in game outcomes. These models can be used to extract estimates of team strength, the between-season, within-season, and game-to-game variability of team strengths, as well each team's home advantage. We implement our approach across a decade of play in each of the National Football League (NFL), National Hockey League (NHL), National Basketball Association (NBA), and Major League Baseball (MLB), finding that the NBA demonstrates both the largest dispersion in talent and the largest home advantage, while the NHL and MLB stand out for their relative randomness in game outcomes. We conclude by proposing new metrics for judging competitiveness across sports leagues, both within the regular season and using traditional postseason tournament formats. Although we focus on sports, we discuss a number of other situations in which our generalizable models might be usefully applied.

\end{abstract}

\begin{keyword}
\kwd{sports analytics}
\kwd{Bayesian modeling}
\kwd{competitive balance}
\kwd{MCMC}
\end{keyword}

\end{frontmatter}


\section{Introduction}
Most observers of sport can agree that game outcomes are to some extent subject to chance. The line drive that miraculously finds the fielder's glove, the fumble that bounces harmlessly out-of-bounds, the puck that ricochets into the net off of an opponent's skate, or the referee's whistle on a clean block can all mean the difference between winning and losing. Yet game outcomes are not \emph{completely} random---there are teams that consistently play better or worse than the average team. To what extent does luck influence our perceptions of team strength over time? 

One way in which statistics can lead this discussion lies in the untangling of signal and noise when comparing the caliber of each league's teams. For example, is team $i$ better than team $j$? And if so, how confident are we in making this claim? Central to such an understanding of sporting outcomes is that if we know each team's relative strength, then, \textit{a priori}, game outcomes---including wins and losses---can be viewed as unobserved realizations of random variables. As a simple example, if the probability that team $i$ beats team $j$ at time $k$ is 0.75, this implies that in a hypothetical infinite number of games between the two teams at time $k$, $i$ wins three times as often as $j$. Unfortunately, in practice, team $i$ will typically only play team $j$ once at time $k$. Thus, game outcomes alone are unlikely to provide enough information to precisely estimate true probabilities, and, in turn, team strengths.

Given both national public interest and an academic curiosity that has extended across disciplines, many innovative techniques have been developed to estimate team strength. These approaches typically blend past game scores with game, team, and player characteristics in a statistical model. Corresponding estimates of talent are often checked or calibrated by comparing out-of-sample estimated probabilities of wins and losses to observed outcomes. Such exercises do more than drive water-cooler conversation as to which team may be better. Indeed, estimating team rankings has driven the development of advanced statistical models \citep{bradley1952rank, glickman1998state} and occasionally played a role in the decision of which teams are eligible for continued postseason play \citep{BCS}.

However, because randomness manifests differently in different sports, a limitation of sport-specific models is that inferences cannot generally be applied to other competitions. As a result,  researchers who hope to contrast one league to another often focus on the one outcome common to all sports: won-loss ratio. Among other flaws, measuring team strength using wins and losses performs poorly in a small sample size, ignores the game's final score (which is known to be more predictive of future performance than won-loss ratio \citep{boulier2003predicting}), and is unduly impacted by, among other sources, fluctuations in league scheduling, injury to key players, and the general advantage of playing at home. In particular, variations in season length between sports---NFL teams play 16 regular season games each year, NHL and NBA teams play 82, while MLB teams play 162---could invalidate direct comparisons of win percentages alone. As an example, the highest annual team winning percentage is roughly 87\% in the NFL but only 61\% in MLB, and part (but not all) of that difference is undoubtedly tied to the shorter NFL regular season. As a result, until now, analysts and fans have never quite been able to quantify inherent differences between sports or sports leagues with respect to randomness and the dispersion and evolution of team strength. We aim to fill this void. 

In the sections that follow, we present a unified and novel framework for the simultaneous comparison of sporting leagues, which we implement to discover inherent differences in North American sport. First, we validate an assumption that game-level probabilities provided by betting markets provide unbiased and low-variance estimates of the true probabilities of wins and losses in each professional contest. Second, we extend Bayesian state-space models for paired comparisons \citep{glickman1998state} to multiple domains. These models use the game-level betting market probabilities to capture implied team strength and variability. Finally, we present unique league-level properties that to this point have been difficult to capture, and we use the estimated posterior distributions of team strengths to propose novel metrics for assessing league parity, both for the regular season and postseason. We find that, on account of both narrower distributions of team strengths and smaller home advantages, a typical contest in the NHL or MLB is much closer to a coin-flip than one in the NBA or NFL.

\subsection{Literature review}

The importance of quantifying team strength in competition extends across disciplines. This includes contrasting league-level characteristics in economics \citep{leeds2004economics}, estimating game-level probabilities in statistics \citep{glickman1998state}, and classifying future game winners in forecasting \citep{boulier2003predicting}. We discuss and synthesize these ideas below.

\subsubsection{Competitive balance}

Assessing the competitive balance of sports leagues is particularly important in economics and management \citep{leeds2004economics}. While competitive balance can purportedly measure several different quantities, in general it refers to levels of equivalence between teams. This could be equivalence within one time frame (e.g. how similar was the distribution of talent within a season?), between time frames (e.g. year-to-year variations in talent), or from the beginning of a time frame until the end (e.g. the likelihood of each team winning a championship at the start of a season).

The most widely accepted within-season competitive balance measure is Noll-Scully \citep{noll1988professional, scully1989business}. It is computed as the ratio of the observed standard deviation in team win totals to the idealized standard deviation, which is defined as that which would have been observed due to chance alone if each team were equal in talent. Larger Noll-Scully values are believed to reflect greater imbalance in team strengths. 

While Noll-Scully has the positive quality of allowing for interpretable cross-sport comparisons, a reliance on won-loss outcomes entails undesireable properties as well \citep{owen2010limitations, owen2015competitive}. For example, Noll-Scully increases, on average, with the number of games played \citep{owen2015competitive}, hindering any comparisons of the NFL (16 games) to MLB (162), for example. Additionally, each of the leagues employ some form of an unbalanced schedule. Teams in each of MLB, the NBA, NFL, and NHL play intradivisional opponents more often than interdivisional ones, and intraconference opponents more often than interconference ones, meaning that one team's won-loss record may not be comparable to another team's due to differences in the respective strengths of their opponents \citep{lenten2015measurement}. Moreover, the NFL structures each season's schedule so that teams play interdivisional games against opponents that finished with the same division rank in the standings in the prior year. In expectation, this punishes teams that finish atop standings with tougher games, potentially driving winning percentages toward 0.500. Unsurprisingly, unbalanced scheduling and interconference play can lead to imprecise competitive balance metrics derived from winning percentages \citep{utt2002pitfalls}. As one final weakness, varying home advantages between sports leagues, as shown in \cite{moskowitz2011scorecasting}, could also impact comparisons of relative team quality that are predicated on wins and losses. 

Although metrics for league-level comparisons have been frequently debated, the importance of competitive balance in sports is more uniformly accepted, in large part due to the uncertainty of outcome hypothesis \citep{rottenberg1956baseball, knowles1992demand, lee2008attendance}. Under this hypothesis, league success---as judged by attendance, engagement, and television revenue---correlates positively with teams having equal chances of winning. Outcome uncertainty is generally considered on a game-level basis, but can also extend to season-level success (i.e, teams having equivalent chances at making the postseason). As a result, it is in each league's best interest to promote some level of \emph{parity}---in short, a narrower distribution of team quality---to maximize revenue \citep{crooker2007sports}. Related, the Hirfindahl-Hirschman Index \citep{owen2007measuring} and Competitive Balance Ratio \citep{humphreys2002alternative} are two metrics attempting to quantify the relative chances of success that teams have within or between certain time frames. 

 \subsubsection{Approaches to estimating team strength}

Competitive balance and outcome uncertainty are rough proxies for understanding the distribution of talent among teams. For example, when two teams of equal talent play a game without a home advantage, outcome uncertainty is maximized; e.g., the outcome of the game is equivalent to a coin flip. These relative comparisons of team strength began in statistics with paired comparison models, which are generally defined as those designed to calibrate the equivalence of two entities. In the case of sports, the entities are teams or individual athletes. 

The Bradley-Terry model (BTM, \cite{bradley1952rank}) is considered to be the first detailed paired comparison model, and the rough equivalent of the soon thereafter developed Elo rankings \citep{elo1978rating, glickman1995comprehensive}. Consider an experiment with $t$ treatment levels, compared in pairs. BTM assumes that there is some true ordering of the probabilities of efficacy, $\pi_{1}, \ldots, \pi_{t}$, with the constraints that $\sum_{i=1}^{t}\pi_{i} = 1$ and $\pi_{i}\geq 0$ for $i = 1,\ldots,t$.  When comparing treatment $i$ to treatment $j$, the probability that treatment $i$ is preferable to $j$ (i.e. a win in a sports setting) is computed as $\frac{\pi_{i}}{\pi_i+\pi_j}$. 

\cite{glickman1998state} and \cite{glickman2016estimating} build on the BTM by allowing team-strength estimates to vary over time through the modeling of point differential in the NFL, which is assumed to follow an approximately normal distribution.  Let $y_{(s,k)ij}$ be the point differential of a game during week $k$ of season $s$ between teams $i$ and $j$. In this specification, $i$ and $j$ take on values between $1$ and $t$, where $t$ is the number of teams in the league. Let $\theta_{(s,k)i}$ and $\theta_{(s,k)j}$ be the strengths of teams $i$ and $j$, respectively, in season $s$ during week $k$, and let $\alpha_i$ be the home advantage parameter for team $i$ for $i = 1,\ldots, t$. \cite{glickman1998state} assume that for a game played at the home of team $i$ during week $k$ in season $s$,  

$$
E[y_{(s,k)ij} | \theta_{(s,k)i},\theta_{(s,k)j},\alpha_{i} ] = \theta_{(s,k)i} - \theta_{(s,k)j} + \alpha_i,
$$

\noindent where $E[y_{(s,k)ij} | \theta_{(s,k)i},\theta_{(s,k)j},\alpha_{i}]$ is the expected point differential given $i$ and $j$'s team strengths and the home advantage of team $i$.

The model of \cite{glickman1998state} allows for team strength parameters to vary stochastically in two distinct ways: from the last week of season $s$ to the first week of season $s+1$, and from week $k$ of season $s$ to week $k+1$ of season $s$. As such, it is termed a `state-space' model, whereby the data is a function of an underlying time-varying process plus additional noise. 

\cite{glickman1998state} propose an autoregressive process to model team strengths, whereby over time, these parameters are pulled toward the league average. Under this specification, past and future season performances are incorporated into season-specific estimates of team quality. Perhaps as a result, \cite{koopmeiners2012comparison} identifies better fits when comparing state-space models to BTM's fit separately within each season. Additionally, unlike BTM's, state-space models would not typically suffer from identifiability problems were a team to win or lose all of its games in a single season (a rare, but extant possibility in the NFL).\footnote{In the NFL, the 2007 New England Patriots won all of their regular season games, while the 2008 Detroit Lions lost all of their regular season games.} For additional and related state-space resources, see \cite{fahrmeir1994dynamic}, \cite{knorr2000dynamic},  \cite{cattelan2013dynamic}, \cite{baker2015time}, and \cite{manner2015modeling}. Additionally, \cite{matthews2005improving}, \cite{owen2011dynamic}, \cite{koopmeiners2012comparison}, \cite{tutz2015extended}, and \cite{wolfson2015s} implement related versions of the original BTM. 

Although the state-space model summarized above appears to work well in the NFL, a few issues arise when extending it to other leagues. First, with point differential as a game-level outcome, parameter estimates would be sensitive to the relative amount of scoring in each sport. Thus, comparisons of the NHL and MLB (where games, on average, are decided by a few goals or runs) to the NBA and NFL (where games, on average, are decided by about 10 points) would require further scaling. Second, a normal model of goal or run differential would be inappropriate in low scoring sports like hockey or baseball, where scoring outcomes follow a Poisson process \citep{mullet1977simeon, thomas2007inter}. Finally, NHL game outcomes would entail an extra complication, as roughly 25\% of regular season games are decided in overtime or a shootout.

In place of paired comparison models, alternative measures for estimating team strength have also been developed. \cite{massey1997statistical} used maximum likelihood estimation and American football outcomes to develop an eponymous rating system. A more general summary of other rating systems for forecasting use is explored by \cite{boulier2003predicting}. In addition, support vector machines and simulation models have been proposed in hockey \citep{demers2015riding, buttrey2016beating}, neural networks and na\"{\i}ve Bayes implemented in basketball \citep{loeffelholz2009predicting, miljkovic2010use}, linear models and probit regressions in football \citep{harville1980predictions, boulier2003predicting}, and two stage Bayesian models in baseball \citep{yang2004two}. While this is a non-exhaustive list, it speaks to the depth and variety of coverage that sports prediction models have generated. 

\subsection{Betting market probabilities}

In many instances, researchers derive estimates of team strength in order to predict game-level probabilities. Betting market information has long been recommended to judge the accuracy of these probabilities \citep{harville1980predictions, stern1991probability}. Before each contest, sports books---including those in Las Vegas and in overseas markets---provide a price for each team, more commonly known as the money line.

Mathematically, if team $i$'s money line is $\ell_i$ against team $j$ (with corresponding money line $\ell_j$), where $|\ell_i| \geq 100$, then the boundary win probability for that team, $p_i(\ell_i)$, is given by:
$$
  p_i(\ell_i) = \begin{cases}
        \frac{100}{100 + \ell_i} & \text{ if } \ell_i \geq 100 \\
        \frac{|\ell_i|}{100 + |\ell_i|} & \text{ if } \ell_i \leq -100
      \end{cases} \,.
$$

The boundary win probability represents the threshold at which point betting on team $i$ would be profitable in the long run.

As an example, suppose the Chicago Cubs were favored ($\ell_i = -127$ on the money line) to beat the Arizona Diamondbacks ($\ell_j = 117$). The boundary win probability for the Cubs would be $p_i(-127) = 0.559$; for the Diamondbacks, $p_j(117) = 0.461$. Boundary win probabilities sum to greater than one by an amount collected by the sportsbook as profit (known colloquially as the ``vig" or ``vigorish").  However, it is straightforward to normalize boundary probabilities to sum to unity to estimate $p_{ij}$, the implied probability of $i$ defeating $j$: 
\begin{eqnarray}
  p_{ij} = \frac{p_i(\ell_i)}{p_i(\ell_i) + p_j(\ell_j)}. \label{eqn:moneyline}
\end{eqnarray}

\noindent In our example, dividing each boundary probability by $1.02 = (0.559 + 0.461)$ implies win probabilities of 54.8\% for the Cubs and 45.2\% for the Diamondbacks. 

In principle, money line prices account for all determinants of game outcomes known to the public prior to the game, including team strength, location, and injuries. Across time and sporting leagues, researchers have identified that it is difficult to estimate win probabilities that are more accurate than the market; i.e, the betting markets are efficient. As an incomplete list, see \cite{harville1980predictions,  gandar1988testing, lacey1990estimation, stern1991probability, carlin1996improved, colquitt2001testing, spann2009sports, nichols2012impact, paul2014market, lopez2015building}. Interestingly, \cite{colquitt2001testing} suggested that the efficiency of college basketball markets was proportional to the amount of pre-game information available---with the amount known about professional sports teams, this would suggest that markets in the NFL, NBA, NHL and MLB are as efficient as they come. \cite{manner2015modeling} merged predictions from a state-space model with those from betting markets, finding that the combination of both predictions only occasionally outperformed betting markets alone.

We are not aware of any published findings that have compared leagues using market probabilities. Given the varying within-sport metrics of judging team quality and the limited between-sport approaches that rely on wins and losses alone, we aim to extend paired comparison models using money line information to better capture relative team equivalence in a method that can be applied generally.

\section{Validation of betting market data}

We begin by confirming the accuracy of betting market data with respect to game outcomes. Regular season game result and betting line data in the four major North American professional sports leagues (MLB, NBA, NFL, and NHL) were obtained for a nominal fee from Sports Insights (\url{https://www.sportsinsights.com}). Although these game results are not official, they are accurate and widely-used. Our models were fit to data from the 2006--2016 seasons, except for the NFL, in which the 2016 season was not yet completed.

These data were more than 99.3\% complete in each league, in the sense that there existed a valid betting line for nearly all games in these four sports across this time period. Betting lines provided by Sports Insights are expressed as payouts, which we subsequently convert into implied probabilities. The average vig in our data set is 1.93\%, but is always positive, resulting in revenue for the sportsbook over a long run of games. In circumstances where more than one betting line was available for a particular game, we included only the line closest to the start time of the game. A summary of our data is shown in Table~\ref{tab:bigfour}.

\begin{table}[ht]
\centering
\begin{tabular}{lr|rr|rr|r}
  \hline
Sport ($q$) & $t_q$ & $n_{games}$ & $\bar{p}_{games}$ & $n_{bets}$ & $\bar{p}_{bets}$ & Coverage \\ 
  \hline
MLB &   30 & 26728 & 0.541 & 26710 & 0.548 & 0.999 \\ 
  NBA &   30 & 13290 & 0.595 & 13245 & 0.615 & 0.997 \\ 
  NFL &   32 & 2560 & 0.563 & 2542 & 0.589 & 0.993 \\ 
  NHL &   30 & 13020 & 0.548 & 12990 & 0.565 & 0.998 \\ 
   \hline
\end{tabular}
\caption{Summary of cross-sport data. $t_q$ is the number of unique teams in each sport $q$. $n_{games}$ records the number of actual games played, while $n_{bets}$ records the number of those games for which we have a betting line. $\bar{p}_{games}$ is the mean observed probability of a win for the home team, while $\bar{p}_{bets}$ is the mean implied probability of a home win based on the betting line. Note that we have near total coverage (betting odds for almost every game) across all four major sports.} 
\label{tab:bigfour}
\end{table}

We also compared the observed probabilities of a home win to the corresponding probabilities implied by our betting market data (Figure~\ref{fig:betting}). In each of the four sports, Hosmer-Lemeshow tests of an efficient market hypothesis using 10 equal-sized bins of games did not show evidence of a lack of fit when comparing the number of observed and expected wins in each bin. Thus, we find no evidence to suggest that the probabilities implied by our betting market data are biased or inaccurate---a conclusion that is supported by the body of academic literature referenced above. Accordingly, we interpret these probabilities as ``true." 

\begin{knitrout}
\definecolor{shadecolor}{rgb}{0.969, 0.969, 0.969}\color{fgcolor}\begin{figure}
\includegraphics[width=\maxwidth]{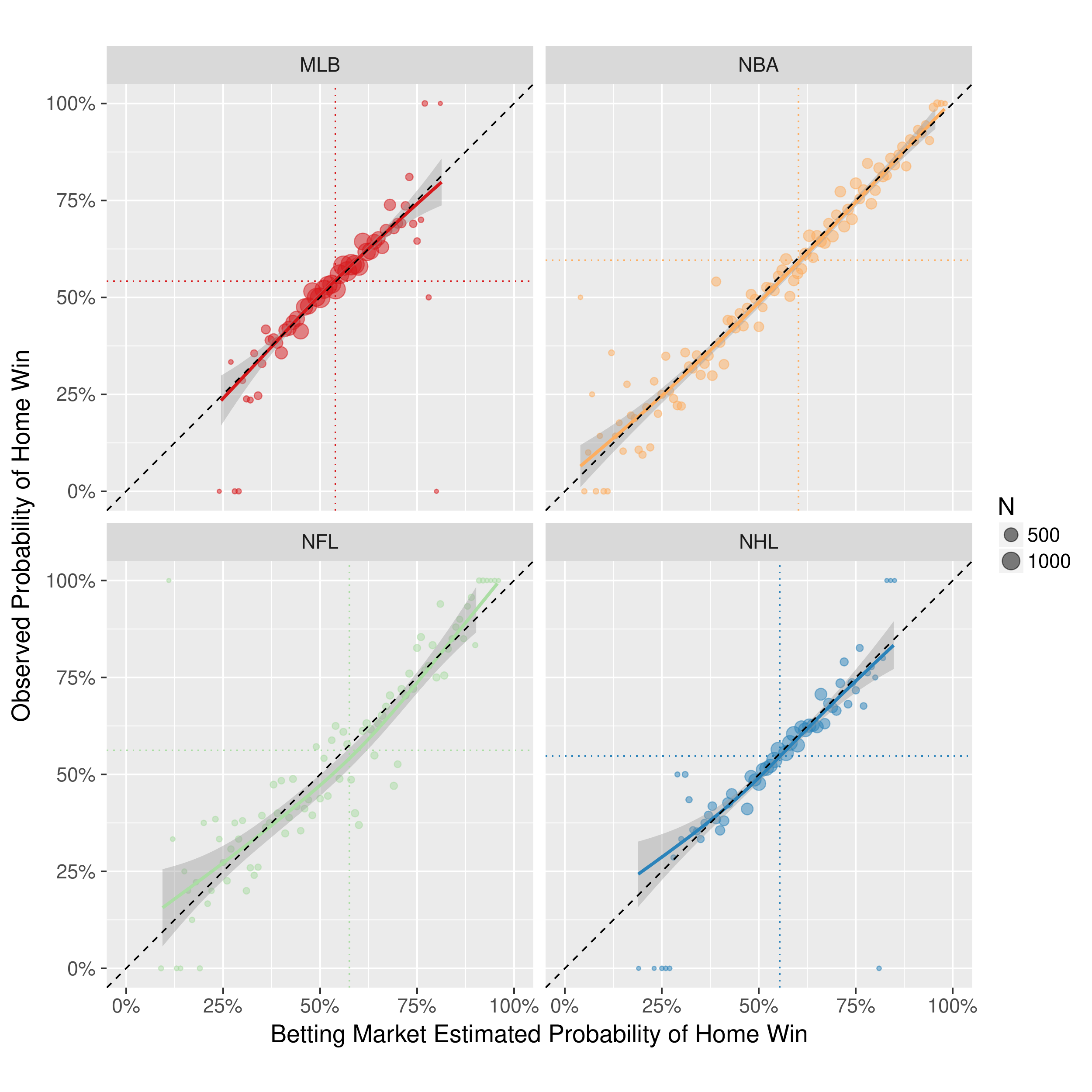} \caption[Accuracy of probabilities implied by betting markets]{Accuracy of probabilities implied by betting markets. Each dot represents a bin of implied probabilities rounded to the nearest hundredth. The size of each dot (N) is proportional to the number of games that lie in that bin. We note that across all four major sports, the observed winning percentages accord with those implied by the betting markets. The dotted diagonal line indicates a completely fair market where probabilities from the betting markets correspond exactly to observed outcomes. In each sport, Hosmer-Lemeshow tests suggest that an efficient market hypothesis cannot be rejected.}\label{fig:betting}
\end{figure}

\end{knitrout}

\section{Bayesian state-space model}

Our model below expands the state-space specification provided by \cite{glickman1998state} to provide a unified framework for contrasting the four major North American sports leagues.

Let $p_{(q,s,k)ij}$ be the probability that team $i$ will beat team $j$ in season $s$ during week $k$ of sports league $q$, for $q \in \{\text{MLB}, \text{NBA}, \text{NFL}, \text{NHL}\}$. The $p_{(q,s,k)ij}$'s are assumed to be known, calculated using sportsbook odds via Equation (\ref{eqn:moneyline}). In using game probabilities, we have a cross-sport outcome that provides more information than only knowing which team won the game or what the score was. 

In our notation, $i,j = 1, \ldots, t_{q}$, where $t_q$ is the number of teams in sport $q$ such that $t_{\text{MLB}} = t_{\text{NBA}} = t_{\text{NHL}} = 30$ and $t_{\text{NFL}} = 32$. Additionally, $s = 1, \ldots, S_q$ and $k = 1, \ldots, K_q$, where $S_q$ and $K_q$ are the number of seasons and weeks, respectively in league $q$. In our data, $K_{\text{NFL}} = 17$, $K_{\text{NBA}} = 25$, $K_{\text{MLB}} = K_{\text{NHL}} = 28$, with $S_{\text{NFL}} = 10$ and $S_{\text{MLB}} = S_{\text{NBA}} = S_{\text{NHL}} = 11$. 

Our next step in building a model specifies the home advantage, and one immediate hurdle is that in addition to having different numbers of teams in each league, certain franchises may relocate from one city to another over time. In our data set, there were two relocations, Seattle to Oklahoma City (NBA, 2008) and Atlanta to Winnipeg (NHL, 2011). Let $\alpha_{q_{0}}$ be the league-wide home advantage (HA) in league $q$, and let $\alpha_{(q) i^{\star}}$ be the team specific effect (positive or negative) for team $i$ among games played in city $i^{\star}$, for $i^{\star} = 1, \ldots,  t^{\star}_{q}$. Here, $t^{\star}_{q}$ is the total number of home cities; in our data, $t^{\star}_{\text{MLB}} = 30$, $t^{\star}_{\text{NBA}} = t^{\star}_{\text{NHL}} = 31$, and $t^{\star}_{\text{NFL}} = 32$.

Letting $\theta_{(q,s,k) i}$ and $\theta_{(q, s, k) j}$ be season-week team strength parameters for teams $i$ and $j$, respectively, we assume that 

$$
E[\text{logit}(p_{(q,s,k) ij}) | \theta_{(q,s,k) i}, \theta_{(q, s, k) j}, \alpha_{q_{0}}, \alpha_{(q) i^{\star}}] = \theta_{(q,s,k) i} - \theta_{(q, s, k) j} + \alpha_{q_0} + \alpha_{(q) i^{\star}},
$$

\noindent where $\text{logit}(.)$ is the log-odds transform. Note that $\theta_{(q,s,k) i}$ and $\theta_{(q,s,k) j}$ reflect absolute measures of team strength, and translate into each team's probability of beating a league average team. We center team strength and individual home advantage estimates about 0 to ensure that our model is identifiable (e.g., $\sum_{i=1}^{t_{q}} \theta_{(q,s,k)i} = 0$ for all $q,s,k$ and $\sum_{i^{\star}=1}^{t^{\star}_{q}} \alpha_{(q) i^{\star}} = 0$ )

Let ${\bf p}_{(q,s,k)}$ represent the vector of length $g_{(q,s,k)}$, the number of games in league $q$ during week $k$ of season $s$, containing all of league $q$'s probabilities in week $k$ of season $s$. Our first model of game outcomes, henceforth referred to as the individual home advantage model (Model IHA), assumes that
\begin{eqnarray}
\text{logit}({\bf p}_{(q,s,k)}) \sim N(\mathbf{\theta}_{(q,s,k)}\mathbf{X}_{(q,s,k)} + \alpha_{q_0}\mathbf{J}_{g_{(q,s,k)}} + \balpha_{q}\mathbf{Z}_{(q,s,k)}, \sigma^{2}_{q,game}\mathbf{I}_{g_{(q,s,k)}}),   \nonumber
\end{eqnarray}

\noindent where $\mathbf{\theta}_{(q,s,k)}$ is a vector of length $t_{q}$ containing the team strength parameters in season $s$ during week $k$ and $\balpha_{q} = \left\{\alpha_{(q) 1}, \ldots, \alpha_{(q) t^{\star}_{q}}    \right\}$. Note that $\balpha_{q}$ does not vary over time (i.e. HA is assumed to be constant for a team over weeks and seasons). $\mathbf{X}_{(q,s,k)}$ and $\mathbf{Z}_{(q,s,k)}$ contain $g_{(q,s,k)}$ rows and $t_{q}$ and $t^{\star}_{q}$ columns, respectively.   The matrix $\mathbf{X}_{(q,s,k)}$ contains the values $\{1, 0, -1\}$ where for a given row (i.e. one game) the value of $i^{th}$ column in that row is a 1/-1 if the $i^{th}$ team played at home/away in the given game and 0 otherwise. $\mathbf{Z}_{(q,s,k)}$ is a matrix containing a 1 in column $i^{\star}$ if the corresponding game was played in city $i^{\star}$, and 0 otherwise. Finally, $\sigma^{2}_{q,game}$ is the game-level variance, $\mathbf{J}_{g_{(q,s,k)}}$ is a column vector of length $g_{(q,s,k)}$ containing all 1's, and $\mathbf{I}_{g_{(q,s,k)}}$ is an identity matrix with dimension ${g_{(q,s,k)}} \times {g_{(q,s,k)}}$.\\

In addition, we propose a simplified version of Model IHA, labelled as Model CHA (constant home advantage), which assumes that the HA within each sport is identical for each franchise, such that
\begin{eqnarray}
\text{logit}({\bf p}_{(q,s,k)}) &\sim& N(\mathbf{\theta}_{(q,s,k)}\mathbf{X}_{(q,s,k)} + \alpha_{q_0}\mathbf{J}_{g_{(q,s,k)}}, \sigma^{2}_{q,game}\mathbf{I}_{g_{(q,s,k)}}). \nonumber
\end{eqnarray}

\noindent In Model CHA, matrices ${\bf p}_{(q,s,k)}$, $\mathbf{X}_{(q,s,k)}$, $\mathbf{J}_{g_{(q,s,k)}}$, and $\mathbf{I}_{g_{(q,s,k)}}$ are specified identically to Model IHA. As a result, for a game between home team $i$ and away team $j$ during week $k$ of season $s$, $E[\text{logit}(p_{(q,s,k) ij})] = \theta_{(q,s,k) i} - \theta_{(q, s, k) j} + \alpha_{q_0}$ under Model CHA. 

Similar to \cite{glickman1998state}, we allow the strength parameters of the teams to vary auto-regressively from season-to-season and from week-to-week. In general, this entails that team strength parameters are shrunk towards the league average over time in expectation. Formally,
$$
\theta_{(q,s+1,1)} |  {\bf \theta}_{q,s,K_q}, \gamma_{q,\text{season}}, \sigma^{2}_{q,\text{season}} \sim N (\gamma_{q, \text{season}}\mathbf{\theta}_{(q,s,K_q)}, \sigma^{2}_{q,\text{season}}\mathbf{I}_{t_{q}}) $$
for all $s \in 1, \ldots, S_q-1$,  and
$$
\theta_{(q,s,k+1)} | \mathbf{\theta}_{(q,s,k)}, \gamma_{q,week}, \sigma^{2}_{q,\text{week}}  \sim N (\gamma_{q, \text{week}}\mathbf{\theta}_{(q,s,k)},\sigma^{2}_{q,\text{week}}\mathbf{I}_{t_{q}})
$$
for all $s \in 1, \ldots, S_q$, $k \in 1, \ldots, K_q-1$.

In this specification, $\gamma_{q,\text{week}}$ is the autoregressive parameter from week-to-week, $\gamma_{q,\text{season}}$ is the autoregressive parameter from season-to-season, and $\mathbf{I}_{t_{q}}$ is the identity matrix of dimension $t_{q} \times t_{q}$.

Given the time-varying nature of our specification, all specifications use a Bayesian approach to obtain model estimates. For sport $q$, the team strength parameters for week $k=1$ and season $s=1$ have a prior distribution of
$$
\theta_{(q,1,1)i} \sim N(0, \sigma^{2}_{q,\text{season}}) \,, \qquad \text{for all } i \in 1, \ldots, t_{q}.
$$

Team specific home advantage parameters have a similar prior, namely, 
$$
\alpha_{(q)i^\star}\sim N(0, \sigma^{2}_{q,\alpha}) \,, \qquad \text{for } i \in 1, \ldots, t^{\star}_{q}.
$$

Finally, letting $\tau^{2}_{q,\text{game}} = 1 / \sigma^{2}_{q,\text{game}}$, $\tau^{2}_{q,\text{season}} = 1/\sigma^{2}_{q,\text{season}}$, $\tau^{2}_{q,\text{week}} = 1/\sigma^{2}_{q,\text{week}}$, and $\tau^{2}_{q,\alpha} = 1/\sigma^{2}_{q,\alpha}$, we assume the following prior distributions \citep{gelman2006prior}: 
\begin{align*}
\tau^{2}_{q,game} &\sim Uniform(0,1000) &\qquad
  \alpha_{q_0} &\sim N(0,10000) \\
\tau^{2}_{q,season} &\sim Uniform(0,1000) &\qquad
  \gamma_{q,season} &\sim Uniform(0,1) \\
\tau^{2}_{q,week} &\sim Uniform(0,1000) &\qquad
  \gamma_{q,week} &\sim Uniform(0,1.5) \\
\tau^{2}_{q,\alpha} &\sim Uniform(0,1000) && \\
\end{align*}

\noindent Note that we cap $\gamma_{q,\text{week}}$ and $\gamma_{q,\text{season}}$ at 1.5 and 1.0, respectively, corresponding to prior beliefs in whether or not team strengths could explode within (unlikely, but feasible) or between (highly unlikely) seasons. 

One of our main interests lies in gauging the game-level equivalence of each league's teams; i.e., how likely was it or will it be for each team to beat other teams? In this respect, we are interested in both looking backwards across time (descriptive) as well as looking forwards (predictive). However, Models IHA and CHA each blend outcomes from weeks prior to, during, and after week $k$ to estimate team strength. While this is ideal for measuring league parity looking backwards, it is less appropriate to make future game predictions. As such, in each $q$ for season $S_q$ (the last season of our data), we fit a series of state-space models using Model IHA, done on a weekly basis (these are termed $sequential$ fits, as opposed to $cumulative$). Formally, for $k = 2, \ldots, K_{q}$ in season $S_q$, we fit Model IHA only on games during $k$ or prior. $Sequential$ fits can be used to provide a sense of the predictive capability of our model.


Posterior distributions of each parameter are estimated using Markov Chain Monte Carlo (MCMC) methods. We use Gibbs sampling via the \pkg{rjags} package \citep{rjags} in the \R \citep{Rcite} statistical computing environment to obtain posterior distributions, separately for each $q$.\footnote{Alternatively, we could have fit one model and pooled information across sports. Given the large between-league differences in structure, we opt against this approach.} Three chains---using 40,000 iterations after a burn-in of 4,000 draws, fit with a thin of 5 ---yield 8,000 posterior samples in each $q$.\footnote{2000 iterations were used for $sequential$ fits with a burn-in of 1000.} Visual inspection of trace plots with parallel chains are used to confirm convergence. To assess the underlying assumptions of Models IHA and CHA, including our use of the logit transform on our probability outcomes, we use posterior predictive distribution checks, as in \cite{gelman2014bayesian}. Comparisons of Models IHA and CHA are made using the Deviance Information Criterion (DIC, \cite{spiegelhalter2002bayesian}) and by examining each model's posterior predictive distribution.

While we are unable to share the exact betting market data due to licensing restrictions, a simplified version of our game-level data, the data wrangling code, Gibbs sampling code, posterior draws, and the code used to obtain posterior estimates and figures are all posted to a GitHub repository, available at \url{https://github.com/bigfour/competitiveness}.

\section{Model Assessment}

We begin by validating and comparing the fits of Models IHA and CHA.

\subsection{Model fit}

Trace plots of $\alpha_{q_0}$, $\gamma_{q, \text{season}}$, $\gamma_{q, \text{week}}$, $\sigma_{q, game}$, $\sigma_{q, season}$, and $\sigma_{q, week}$ are shown for each $q$ in Figures \ref{fig:MLBtrace}--\ref{fig:NHLtrace} in the Appendix. Visual inspection of these plots does not provide evidence of a lack of convergence or of autocorrelation between draws. These trace plots stem from Model IHA; conclusions are similar when plotting draws from Model CHA. 

Table~\ref{tab:DIC} shows the deviance information criterion (DIC) for each fit in each league, along with the difference in DIC values and the associated standard error (SE). In each of the NHL, NBA, and NFL, fits with a team-specific HA (Model IHA) yielded lower DIC's (lower is better) by a statistically meaningful margin, with the most noticeable difference in fit improvement in the NBA. DIC's were also lower in MLB using Model IHA, although differences were not significant. 

\begin{table}[!ht]
\makebox[\linewidth]{
\begin{tabular}{l r r r}
\hline
 & Model IHA & Model CHA & Difference (SE) \\ \hline
MLB &  -8538 & -8481  & -56.8 (37.9) \\
NBA & 6864  & 7188 & -323.9 (40.5) \\
NFL & 1135  & 1288 & -153.2 (24.3) \\
NHL & -18294  & -18128 & -165.8 (37.7)\\
\hline
\end{tabular}
}
\caption{Deviance information criterion (DIC) by sport and model, along with the difference in DIC and the associated standard errors (SE, in parentheses). IHA: individual home advantage, CHA: constant home advantage \label{tab:DIC}}
\end{table}

These results suggest that chance alone likely does not account for observed differences in the home advantage among teams in the NBA, NHL, and NFL. For the NFL, this implication matches the findings of \cite{glickman1998state}, who identified meaningful between-franchise differences in terms of playing at home. For consistency, results that follow use model estimates from Model IHA. 

\subsection{Posterior predictive checks}

We next address the fit of Models IHA and CHA by looking at the posterior predictive distribution of each. Formally, we assess whether Models IHA and CHA can use draws from their respective posterior distributions to generate game-level data that roughly matches the observed data.

Our specific interest lies in the posterior predictive distribution of the logit of implied probabilities, $p(\text{logit}({\bf \widetilde{p}}_{(q,s,k)})|\text{logit}({\bf p}_{(q,s,k)}))$.  To draw values, we randomly sample from the joint posterior distribution of the parameters (i.e. team strength, home field advantage, and variance parameters).  Then, conditional on the drawn parameters, we randomly draw from the distribution of $\text{logit}({\bf \widetilde{p}}_{(q,s,k)})$.  Recall that in the IHA model, this distribution is assumed to be normal with the following form:

\begin{eqnarray}
\text{logit}({\bf p}_{(q,s,k)}) \sim N(\mathbf{\theta}_{(q,s,k)}\mathbf{X}_{(q,s,k)} + \alpha_{q_0}\mathbf{J}_{g_{(q,s,k)}} + \balpha_{q}\mathbf{Z}_{(q,s,k)}, \sigma^{2}_{q,game}\mathbf{I}_{g_{(q,s,k)}}).   \nonumber
\end{eqnarray}

\noindent We used 20 simulated sets of logit probabilities from this posterior distribution, as well as 20 more from the posterior distribution of Model CHA.

\begin{figure}[h]
\includegraphics[]{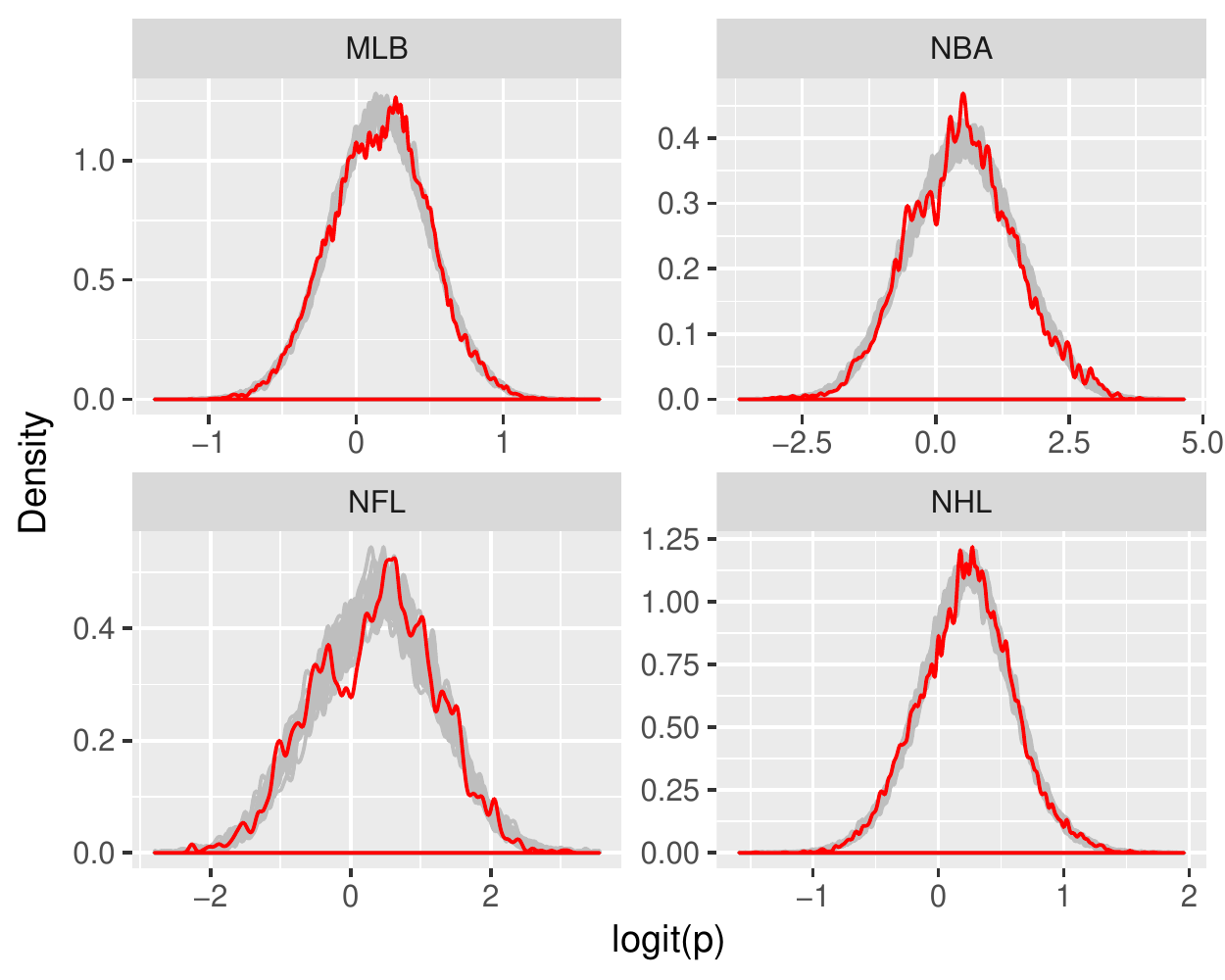}
\caption{Posterior predictive distributions. Density curves of 20 posterior predictive distributions of logit probabilities (in gray) and one curve with the observed distribution of logit probabilities (in red) are overlaid. The bandwith of the density curves is lowered to highlight the jagged nature of sportsbook prices. By and large, the posterior predictive distributions match the observed data. \label{fig:ppd-all}}
\end{figure}

Figure~\ref{fig:ppd-all} overlays each of Model IHA's 20 posterior predictive distributions of logit probabilities (shown in gray density curves) along with the observed distribution of logit probabilities (shown in red).  By and large, the observed distributions of logit probabilities are similar to the simulated distributions in each sport. In particular, the density in the tails of the posterior predictive distributions (reflecting probabilities near 0 or 1) does not show any meaningful departure from the observed distributions.

We purposefully use a lower bandwith for the density curves in Figure~\ref{fig:ppd-all} to highlight interesting discrepancies between the observed and predictive distributions. In the NBA and NFL, for example, the observed distribution is slightly lower than the simulated distributions with logit probabilities near 0 (i.e., both teams have a win probability of 0.5). This is likely occurring due to preference of sportsbooks to set prices that are rounded to the nearest 5 (e.g. -105, -110, -155, etc.). As an example, there are 33 NFL games where the home team's boundary price is -185 (1.3\% of games), and there are 22 other prices that are observed for the home team in 15 or more unique games. Given that Models CHA and IHA do not extract back to rounded prices for each team, it is not surprising that our posterior predictive distributions are smoother than the observed data. Similarly, \cite{glickman1998state} found discrepancies between the observed distribution of point differential in the NFL and the posterior predictive distributions of point differential, on account of the increased likelihood of games ending with margins of victory of 3 or 7 in the NFL.  We believe that we are observing a similar phenomenon, but based on the increased likelihood of a sportsbook to assign rounded odds.    

Next, we use posterior predictive distributions to compare the appropriateness of Models IHA and CHA for each team, as well as to contrast each of the two models to one another. To do this, we calculate the average discrepancy between the mean posterior predictive distribution of each game and the observed game probability, averaged over home team for each model. These team level results are shown in Figure~\ref{fig:ppd-model}. Discrepencies from Model CHA are shown in via circles, with arrows pointing towards the average discrepency for Model IHA. The color of the arrow (blue for yes, red for no) identifies whether, on average, Model IHA more closely matched the observed data than Model CHA. The dashed black line in each plot at 0 on the $x$-axis corresponds to home teams for whom, on average, the mean of the posterior predictive distribution matched that shown in our observed data. 

For 80\% of the teams across all leagues, the posterior predictive distribution using Model IHA more appropriately reflects the observed data. In MLB, the two models perform nearly the same with the exception of the Colorado Rockies, whose home field advantage is underestimated when using Model CHA (see Section~\ref{sec:home}).  Discrepencies in Model IHA offer a slight improvement over those from Model CHA in both the NFL and NHL, with a marked improvement noticed in the NBA. For example, observed home probabilities for Denver, Utah, and Golden State are underestimated using Model CHA, while those for Brooklyn, Detroit, New York, and Philadelphia, are, on average, overestimated. In the NHL, the posterior predictive distribution using Model IHA more closely matches the observed data for 25 of the 30 teams.

\begin{knitrout}
\definecolor{shadecolor}{rgb}{0.969, 0.969, 0.969}\color{fgcolor}\begin{figure}
\includegraphics[width=\maxwidth]{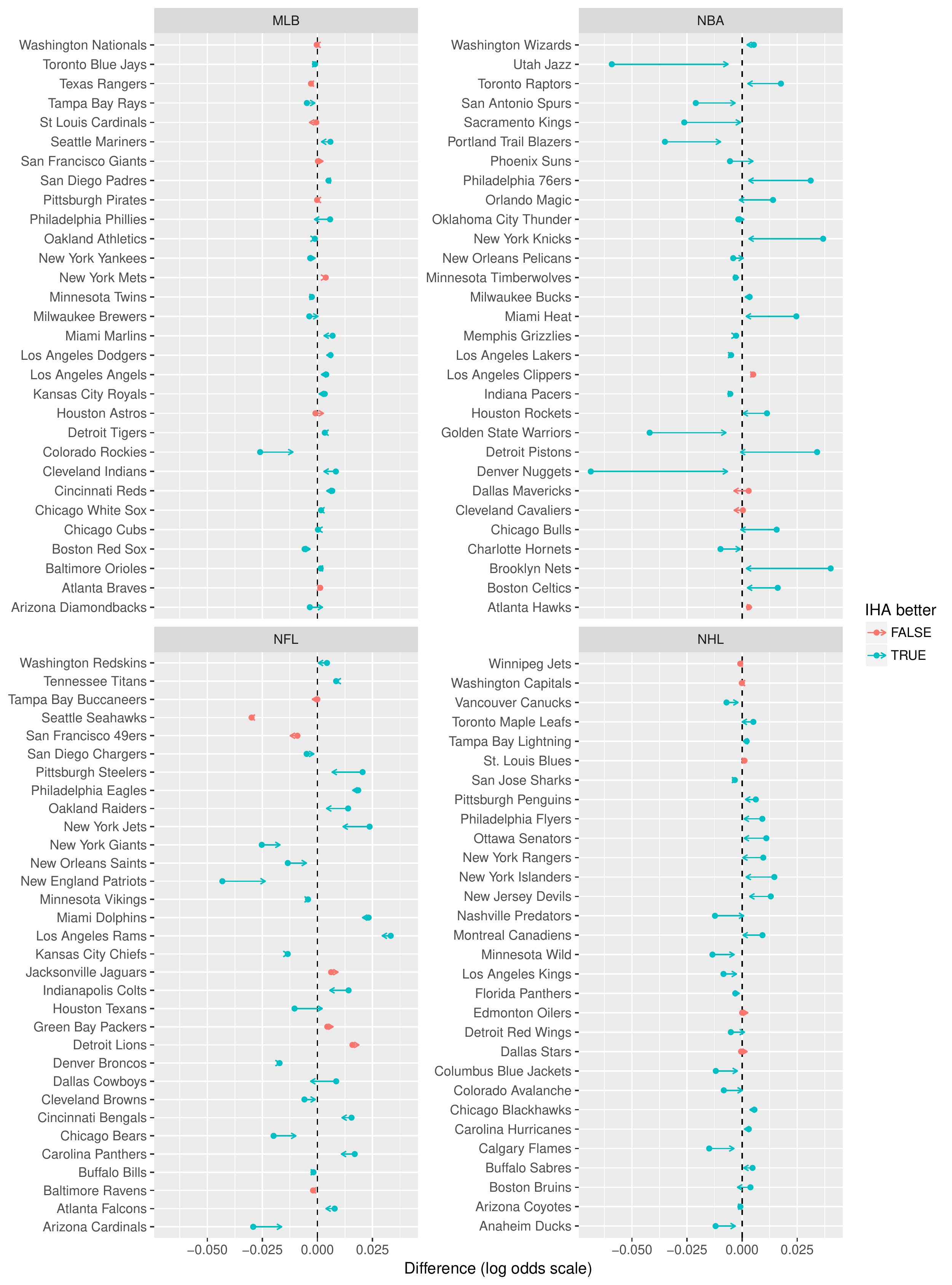} \caption[Posterior predictive distributions by model type]{Posterior predictive distributions by model type. Each dot represents the average difference between the posterior predictive distribution and the truth for each team's home games under the CHA model. The tip of the corresponding arrow represents the same quantity under the IHA model. The difference is smaller under IHA for 80\% of the teams. }\label{fig:ppd-model}
\end{figure}

\end{knitrout}

\section{Results}

In this section we present our results. We discuss the implications of our estimates of team strength and home advantage, as well as the interpretation of our variance and autoregressive parameters. We conclude by evaluating our team strength parameters and illustrating how they could be used for predictive purposes and to build league parity metrics.

\subsection{Team strength}

Table~\ref{tab:thetas} shows summary statistics of the team strength estimates, approximated using posterior mean draws for all weeks $k$ and seasons $s$ across all four sports leagues. Overall, there tends to be a larger variability in team strength at any given point in time in both the NFL and NBA, with average posterior coefficient estimates tending to vary between -1.3 and 1.2 in the NBA and -1.0 and 1.0 in the NFL (on the logit scale) about 95\% of the time. For reference, a team-strength of 1.0 on the log-odds scale implies a $\frac{e^{1.0}}{1+e^{1.0}} = 73.1$\% chance of beating a league average team in a game played at a neutral site. The standard deviation of team strength is smallest in MLB, suggesting that---relative to the other leagues---team strength is more tightly packed. Relative to MLB, spread of team strengths are about 1.3, 3.1, and 3.6 times wider in the NHL, NFL, and NBA, respectively.

\begin{table}[ht]
\centering
\begin{tabular}{lrrrrrrrr}
  \hline
League ($q$) & N* & min & $2.5^{th}$ & Q1 & Q3 & $97.5^{th}$ & max & sd \\ 
  \hline
MLB & 9240 & -0.553 & -0.373 & -0.134 & 0.126 & 0.337 & 0.473 & 0.182 \\ 
  NBA & 8250 & -2.202 & -1.268 & -0.487 & 0.477 & 1.204 & 1.873 & 0.660 \\ 
  NFL & 5440 & -1.576 & -1.092 & -0.402 & 0.416 & 1.030 & 1.906 & 0.559 \\ 
  NHL & 9240 & -1.034 & -0.523 & -0.162 & 0.180 & 0.438 & 0.877 & 0.246 \\ 
   \hline
\end{tabular}
\caption{Summary of average week-level team strength parameters, taken on the log-odds scale. N*: number of unique team strength draws (teams $\times$ seasons $\times$ weeks)} 
\label{tab:thetas}
\end{table}

Figure~\ref{fig:spaghetti} shows estimated team strength coefficients over time. Figures~\ref{fig:spaghetti-mlb}--\ref{fig:spaghetti-nhl} (shown in the Appendix) provide an individual plot for each sport, which include divisional facets to allow easier identification of individual teams. Teams in Figures~\ref{fig:spaghetti} and \ref{fig:spaghetti-mlb}--\ref{fig:spaghetti-nhl} are depicted using their two primary colors, scraped from \url{http://jim-nielsen.com/teamcolors/} via the \pkg{teamcolors} package~\citep{teamcolors} in \R. A color key for all teams appears in Figure~\ref{fig:teamcolors}.

\begin{knitrout}
\definecolor{shadecolor}{rgb}{0.969, 0.969, 0.969}\color{fgcolor}\begin{figure}
\includegraphics[width=\maxwidth]{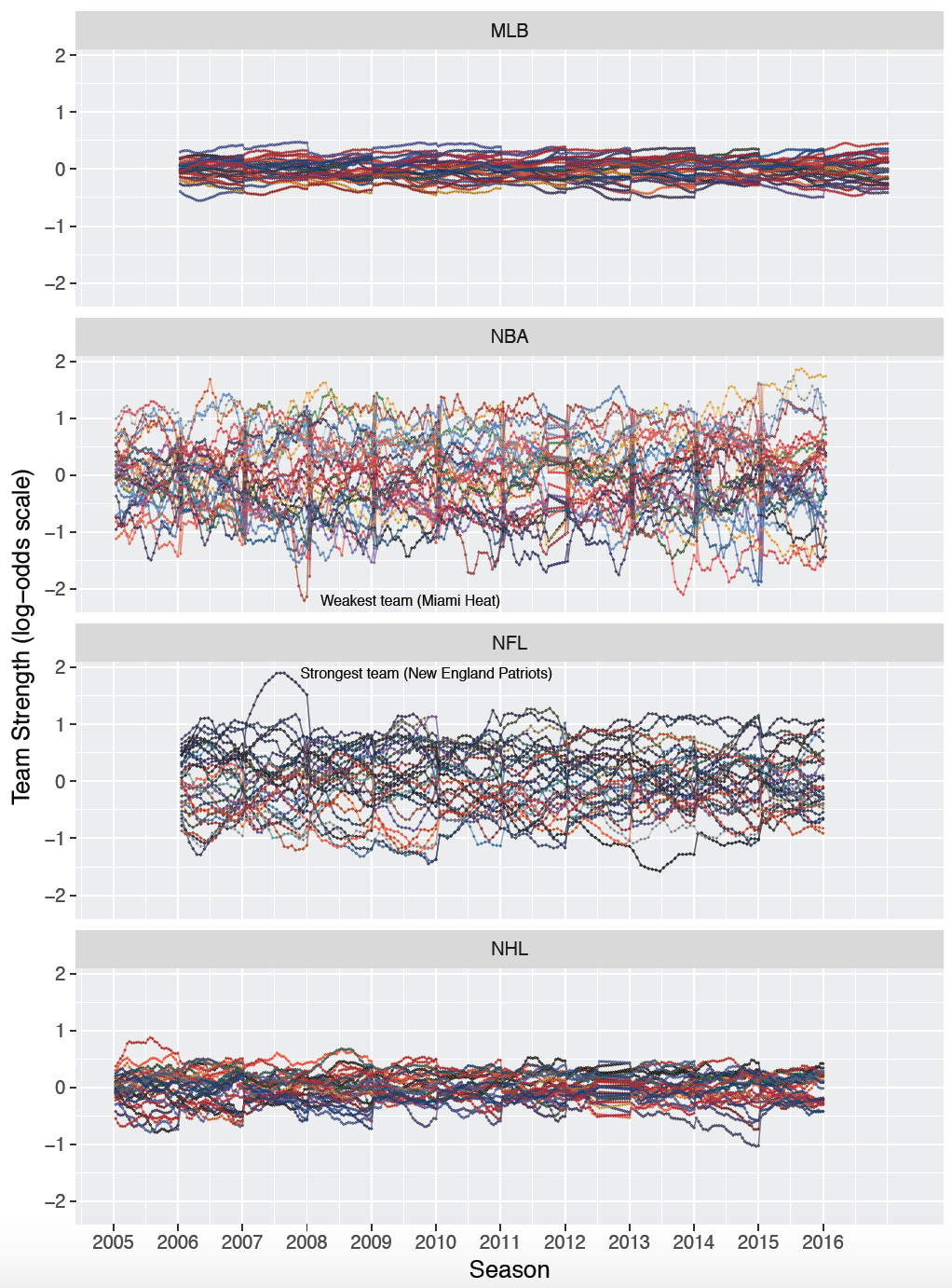} \caption[Mean team strength parameters over time for all four sports leagues]{Mean team strength parameters over time for all four sports leagues. MLB and NFL seasons follow each yearly tick mark on the $x$-axis, while NBA and NHL seasons begin during years labeled by the preceding tick marks.}\label{fig:spaghetti}
\end{figure}

\end{knitrout}

As in Table~\ref{tab:thetas}, these figures suggest that the NBA and NFL boast larger between-team gaps in quality than the NHL and MLB, implying more competitive balance in the latter pair of leagues. On one level, this stands somewhat in contrast to competitive balance as measured using Noll-Scully, which alternatively argues that the NFL is more competitively balanced than MLB~\citep{WagesWins}. One likely explanation for this difference is Null-Scully's link to number of games played, which artificially makes MLB (162 games) appear less balanced than it actually is and the NFL (16) appear more balanced. Like Noll-Scully, we conclude that the NBA shows less competitive balance relative to other leagues.

Our figures also illustrate several other observations. For example, the 2007 New England Patriots of the NFL stand out as having the highest probabilities of beating a league average team, with an average team strength of 1.91 on the log-odds scale, observed during Week 11.  In that season, New England finished the regular season 16-0 before eventually losing in the Super Bowl. The team with the lowest probability of beating a league average team is the NBA's 2007--08 Miami Heat, who during week 23 had a posterior mean team strength of -2.2.  That Heat team finished with an overall record of 15-67, at one point losing 15 consecutive games. Related, it is interesting that the team strength estimates of bad teams in the NBA (e.g. the Heat in 2007--08) lie further from 0 than the estimates for good teams. This possibly reveals the tendency for teams in this league to ``tank"---a strategy of fielding a weak team intentionally to improve the chances of having better selection preference in the upcoming player draft \citep{soebbing2013gamblers}.

Another observation is that in the NHL, top teams appear less dominant than a decade ago. For example, there are seven NHL team-seasons in which at least one team reached an average posterior strength estimate of 0.55 or greater; each of these came during or prior to the 2008--09 season.  In addition to increased parity, the league's point system change in 2005--06---which unintentionally encouraged teams to play more overtime games \citep{lopez2013inefficiencies}---could be responsible. More overtime contests could lead to different perceptions in how betting markets view team strengths, as overtime sessions and the resulting shootouts are roughly equivalent to coin flips \citep{lopez2016predicting}.

As a final point of clarification in Figures~\ref{fig:spaghetti}, \ref{fig:spaghetti-nba}, and \ref{fig:spaghetti-nhl}, the periods of time with straight lines of team strength estimates during the 2012--13 season (NHL) and 2011--12 season (NBA) reflect time lost due to lockouts.

\subsection{Variance and autoregressive parameters}

Table \ref{tab:params} shows the mean and standard deviation of posterior draws for $\gamma_{q, \text{season}}$, $\gamma_{q, \text{week}}$, $\sigma_{q, \text{game}}$, $\sigma_{q, \text{season}}$, and $\sigma_{q, \text{week}}$ for each $q$. Before discussing results from these posterior distributions, it is important to recognize that each variance and autoregressive parameter is uniquely tied to each sport's relative logit scale. For example, the average posterior draw of $\gamma_{NBA, season}$ and $\gamma_{MLB, season}$ are both equal to 0.62, implying that relative to each league's distribution of team strengths, we can expect the same amount of reversion from one season to the next. However, given that there are larger gaps in the team strengths in the NBA, this corresponds to larger reversions in season-level strength when considered on an absolute scale.

\begin{table}[ht]
\centering
\begin{tabular}{llllll}
  \hline
League ($q$) & $\gamma_{q, season}$ & $\gamma_{q, week}$ & $\sigma_{q, game}$ & $\sigma_{q, season}$ & $\sigma_{q, week}$ \\ 
  \hline
MLB & 0.618 (0.031) & 1.002 (0.002) & 0.201 (0.001) & 0.093 (0.005) & 0.027 (0.001) \\ 
  NBA & 0.618 (0.04) & 0.977 (0.003) & 0.274 (0.002) & 0.44 (0.02) & 0.166 (0.003) \\ 
  NFL & 0.69 (0.042) & 0.978 (0.005) & 0.233 (0.008) & 0.331 (0.019) & 0.147 (0.006) \\ 
  NHL & 0.542 (0.027) & 0.993 (0.003) & 0.105 (0.001) & 0.121 (0.006) & 0.053 (0.001) \\ 
   \hline
\end{tabular}
\caption{Mean posterior draw (standard deviation) by league.} 
\label{tab:params}
\end{table}

Posterior draws of $\sigma_{q, \text{game}}$ suggest that the highest game-level errors in our log-odds probability estimates occur in the NBA (median posterior draw of $\sigma_{NBA, game}$ = 0.274), followed in order by the NFL, MLB, and the NHL. Interestingly, although Figure~\ref{fig:spaghetti} identifies that the talent gap between teams is smallest in MLB, $\sigma_{\text{MLB}, \text{game}} \approx 2 \times \sigma_{\text{NHL}, \text{game}}$ in our posterior draws. We posit that this additional game-level error in MLB is a function of the league's pitching match-ups, in which teams rotate through a handful of starting pitchers of varying calibers. 

We also examine the joint distribution of the variability in team strength on a season-to-season ($\sigma_{q, \text{season}}$) and week-to-week ($\sigma_{q, \text{week}}$) basis via the contour plot in Figure~\ref{fig:contourSigma} (Appendix), using separate colors for each $q$. Figure~\ref{fig:contourSigma} reveals that the highest uncertainty with respect to team strength occurs in the NBA, followed in order by the NFL, NHL, and MLB. 

Even when accounting for the larger scale in outcomes, the NBA still stands out as far as increased between-week uncertainty. There are a few possible explanations for this. Injuries, the resting of starters, and in-season trades would seemingly have a larger impact in a sport like basketball where fewer players are participating at a single point in time. In particular, our model cannot precisely gauge team strength when star players who could play are rested in favor of inferior players. Relative to the other professional leagues, star players take on a more important role in the NBA \citep{berri2006road}, an observation undoubtedly known in betting markets. That said, while there is increased variability in our estimate of NBA team strengths, when considering differences in team talent to begin with, these absolute differences are not as extreme (e.g., a difference in team strength of 0.05 means less in the NBA as far as relative ranking than in the NHL).

Figure~\ref{fig:contourGamma} (Appendix) displays the joint posterior distribution of $\gamma_{q, \text{season}}$ and $\gamma_{q, \text{week}}$ via contour plots for each $q$. On a season-to-season basis, team strengths in each of the leagues tend to revert towards the league average of zero as all draws of $\gamma_{q, \text{season}} < 1$ for all $q$. Reversion towards the mean is largest in the NHL (estimated $\gamma_{NHL, \text{season}}$ = 0.54, implying 46\% reversion), followed by the NBA (38\%), MLB (38\% reversion), and the NFL (31\%). However, the only pair of leagues with non-overlapping credible intervals are the NFL and NHL. Note that one reason that team strengths may revert towards zero each year is the structure of each league's draft, in which newly eligble players are chosen. In expectation, the worst team in each league is most likely to get the top selection in the following year's draft, and so by aquiring the best perceived talent, those worst teams are more likely to improve. Perhaps one reason that the NFL shows the most consistency over time is that, in general, it is the worst at drafting newly eligible players (see~\cite{lopez2016draft} for comparisons in the drafting ability of each league).

For each of the NHL, NBA, and NFL, posterior estimates of $\gamma_{q, \text{week}}$ (as well as 95\% credible intervals) imply an autoregressive nature to team strength within each season. Interestingly, the NBA and NFL are the least consistent leagues on a week-to-week basis. In MLB, however, team strength estimates quite possibly follow a random walk (i.e., $\gamma_{\text{MLB}, \text{week}} = 1$), in which the succession of team strength is unpredictable. Alternatively, it is also feasible that MLB team strengths could explode over time ($\gamma_{\text{MLB}, \text{week}} > 1$), in which case these estimates would be pulled towards 0 in the long run (across seasons, via $\gamma_{\text{MLB}, \text{season}})$.

Finally, it is worth noting that our estimates for $\gamma_{\text{NFL}, \text{week}}$ and $\gamma_{\text{NFL}, \text{season}}$---0.98 and 0.69, respectively---do not substantially diverge from the estimates observed by \cite{glickman1998state} (0.99 and 0.82). Further, our credible intervals are narrower. For example, our 95\% credible interval for $\gamma_{\text{NFL}, \text{season}}$ of (0.61, 0.77) is entirely contained within the interval of $(0.52, 1.28)$ reported by \cite{glickman1998state}. In fairness, it is unclear if the decreased uncertainty is a function of our model specification (using log-odds of the probability of a win as the outcome, as opposed to point differential) or because we used a larger sample (10 seasons, compared to 5). 

Like \cite{glickman1998state}, we also observe an inverse link in posterior draws of $\gamma_{\text{NFL}, \text{week}}$ and $\gamma_{\text{NFL}, \text{season}}$. Given that total shrinkage across time is the composite of within- and between-season shrinkage, such an association is not surprising~\citep{glickman1998state}. If one source of reversion towards the average were to increase, the other would likely compensate by decreasing. 

\subsection{The home advantage}
\label{sec:home}

Figure~\ref{fig:alphaAll} shows the 2.5th percentile, median, and 97.5th percentile draws of each team's estimated home advantage parameter, presented on the probability scale. These are calculated by summing draws of $\alpha_{q_0}$ and $\alpha_{(q)i^\star}$ for all $i^\star$. HAs are shown in descending order to provide a sense of the magnitude of differences between the home advantage provided in MLB (league-wide, a 54.0\% probability of beating a team of equal strength at home), NHL (55.5\%), NFL (58.9\%), and NBA (62.0\%). The two franchises that have relocated in the last decade, the Atlanta Thrashers (NHL) and Seattle Supersonics (NBA), are also included for the games played in those respective cities.

\begin{knitrout}
\definecolor{shadecolor}{rgb}{0.969, 0.969, 0.969}\color{fgcolor}\begin{figure}
\includegraphics[width=\maxwidth]{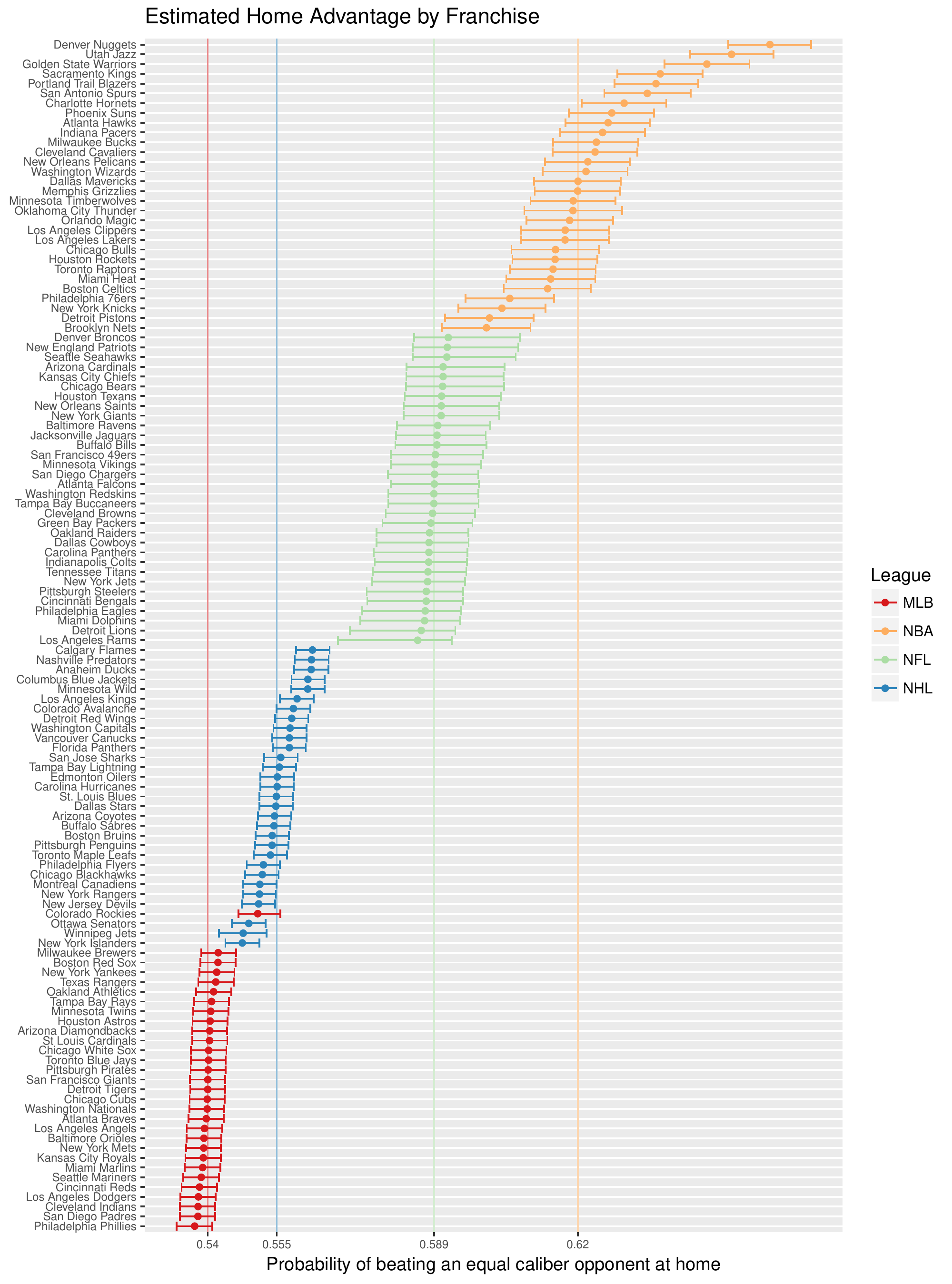} \caption[Median posterior draw (with 2.5th, 97.5th quantiles) of each franchise's home advantage intercept, on the probability scale]{Median posterior draw (with 2.5th, 97.5th quantiles) of each franchise's home advantage intercept, on the probability scale. We note that the magnitude of home advantages are strongly segregated by sport, with only one exception (the Colorado Rockies). We also note that no NFL team, nor any MLB team other than the Rockies, has a home advantage whose 95\% credible interval does not contain the league median. }\label{fig:alphaAll}
\end{figure}

\end{knitrout}

Figure~\ref{fig:alphaAll} depicts substantial between-franchise differences in the home advantage within both the NBA and NHL, with lesser between-franchise differences in MLB and the NFL.

Interestingly, the draws of the home advantage parameters for of a few NFL franchises are skewed (see Denver and Seattle, relative to Detroit), potentially the result of a shorter regular season. Alternatively, the NFL's HA may vary by season, game time, or the day of the game. Anecdotally, night games (Thursday, Sunday, or Monday) conceivably offer a larger HA than those played during the day \citep{Seahawks}. Informally, NFL team-level HA estimates are similar in effect size to those depicted by \cite{koopmeiners2012comparison}.

In the NBA, Denver (first) and Utah (second) post the best home advantages, with Brooklyn showing the worst. This matches the results of \cite{Jazz}, who found significantly better performances when comparing Denver and Utah to the rest of the league with respect to home and road point differential. In MLB, the Colorado Rockies stand out for having the highest home advantage, while the remaining 29 teams boast overlapping credible intervals. We note that teams playing at home in Denver have the largest home advantages in MLB, the NBA, and the NFL, and the 7th-highest in the NHL. We speculate that this consistent advantage across sports is related to the home team's acclimation to the city's notably high altitude.  

Differences between teams within the NBA have plausible impacts on league standings. An NBA team with a typical home advantage can expect to win 62.0\% of home games against a like-caliber opponent. Yet for Brooklyn, the corresponding figure is 60\%, while for Denver, it is 66.1\%. Across 41 games (the number each team plays at home), this implies that Denver's home advantage is worth an extra 1.68 wins in a single season, relative to a league average team. Compared to Brooklyn, Denver's home advantage is worth an estimated 2.5 wins per year. As one important caveat, our model estimates do not account for varying line-up and injury information. If opposing teams were to rest their star players at Denver, for example, our model would artificially inflate Denver's home advantage. 

As a final note, it is interesting that in comparing leagues, the relative magnitudes of the home advantage match the relative standard deviations in team strength (with the NBA the highest, followed in order by NFL, NHL, MLB). To check whether or not the home advantage parameters are independent of team strength estimates (as implied in our model specification), we compared the average posterior draw of the home advantage versus the average posterior team strength across all weeks and seasons for each franchise in each sport (plot not shown). Within each sport, there was no obvious link between average team quality and that team's home intercept, as assessed using scatter plots with a LOESS regression line. That said, further research may be needed to precisely define home advantage in light of varying team stregnth estimates, as well game-level characteristics such as time (i.e., afternoon, night) and day (i.e., weekend, weekday.)

\subsection{Evaluation of team strength estimates}

Ultimately, estimates from Model IHA are designed to estimate team quality at any given point in a season while accounting for factors such as the home advantage and opponent caliber. If these estimates more properly assess team quality than traditional metrics (e.g., won-loss percentage or point differential), they should more accurately link to future performance, such as how well teams will perform over the remainder of the season. Additionally, game-level probabilities estimated from our team strength coefficients should closely track the observed money lines.

That said, it is admittedly unfair to use $cumulative$ estimates of team strength to predict past game outcomes, as future information is implicity used to inform those same game outcomes. In this sense, $sequential$ fits are more appropriate for understanding the predictive capability of our state-space models. 

Figure~\ref{fig:R2} shows the coefficient of determination ($R^2$) between each team's future won-loss percentage in a season and each team's (i) average team strength estimates from $sequential$ Model IHA's, (ii) season-to-date cumulative point differential, and (iii) season-to-date won-loss percentage. Within each sport, this is computed by game number, which helps to account for league-level differences in season length. For purposes of using $sequential$ team strength estimates, we used the mean posterior draw from fits that ended the week prior.

\begin{knitrout}
\definecolor{shadecolor}{rgb}{0.969, 0.969, 0.969}\color{fgcolor}\begin{figure}
\includegraphics[width=\maxwidth]{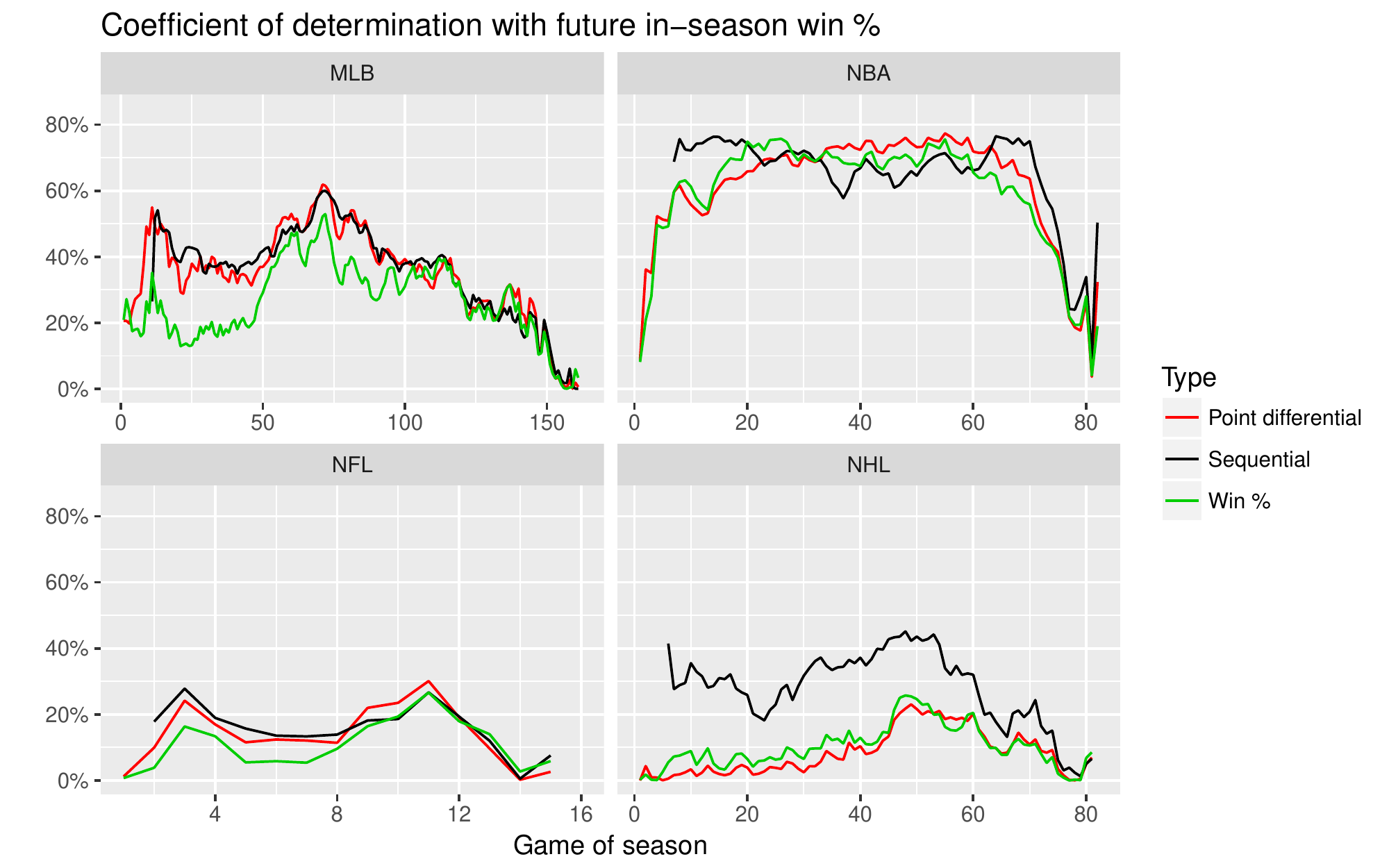} \caption[Coefficient of determination with future in-season win percentage]{Coefficient of determination with future in-season win percentage. We note the improvement our team strength estimates offer over season-to-date win percentage and season-to-date point differential in most sports, especially early in the season. $R^2$ values tend to 0 as the number of future games goes to 0. \ }\label{fig:R2}
\end{figure}

\end{knitrout}

Across each sport, our estimates of team strength generally outperform past team win percentage and point differential in predicting future win percentage. This gap is most pronounced earlier in each season, which is not surprising given the instability of won-loss percentage and point differential in a small number of games. Differences remain throughout most of the regular season in MLB, the NHL, and the NFL. However, by the NBA's mid-season, won-loss ratio and point differential are similar to our estimates of team strength in assessing future performance. By and large, this confirms the findings of \cite{wolfson2015s}, who identified that most of the information needed to predict the remainder of the NBA season is contained within the first third of the year. 

As a second check of predictive accuracy, we compare these predicted game-level probabilities to known game outcomes. Table \ref{tab:ll} highlights the area under the receiver operating characteristic curve (AUC), which calculates the expectation that a randomly drawn probability from a winning home team is greater than a randomly drawn probability of a losing home team (higher is better). Also included is the Brier score (lower is better), along with an accompanying p-value as implemented for calibration accuracy in \cite{spiegelhalter1986probabilistic}.

\begin{table}[ht]
\centering
\begin{tabular}{|c|cc|cc|}
  \hline
   & \multicolumn{2}{c|}{AUC} & \multicolumn{2}{c|}{Brier Score} \\
  League ($q$) & observed & sequential & observed (p-value) & sequential (p-value)  \\
 \hline
MLB & 0.605 & 0.573 & 0.241 (0.996) & 0.245 (0.333) \\ 
  NBA & 0.756 & 0.756 & 0.194 (0.803) & 0.194 (0.759) \\ 
  NFL & 0.682 & 0.685 & 0.226 (0.34) & 0.226 (0.548) \\ 
  NHL & 0.595 & 0.589 & 0.242 (0.88) & 0.243 (0.486) \\ 
   \hline
\end{tabular}
\caption{AUC values and Brier scores (p-values) by sport. Observed probabilities use known probabilities from betting markets, while sequential probabiltiies use predictions from posterior draws using sequential fits of Model IHA.} 
\label{tab:ll}
\end{table}

For each of the NBA, NFL, and NHL, AUC and Brier metrics suggest that predictions made from sequential fits can closely approximate the observed game probabilities. However, our predictions yield a lower AUC and a higher Brier score in MLB, which likely reflects our inability to account for each game's starting pitcher. 

Although results from these predictions do not suggest an existance of an arbitrage opportunity (recall that sports books add a vig to each team's price), they do imply that both our team strength and home advantage estimates can be used to extract accurate game-level projections. Further, that there is no major deviation from the observed data is comforting with respect to our choice of a model for the game probabilities. 

\subsection{How often does the best team win?  A new measure of league parity}

We conclude by addressing our initial question about the inherent randomness of game outcomes.\footnote{Our approach here is not unlike that of \cite{james1993answering}.}  

One simple way to compare league randomness would be to contrast the observed distribution of $p_{(q,s,k) ij}$'s between each $q$. However, while sportsbook odds can be used to infer the probability of each team winning, these odds are only provided for scheduled games. As a result, any between-league comparisons using sportsbook odds alone would be contingent upon each league's actual schedule, and they may not accurately reflect differences that would be observed if all teams were to play one another.

A second option would be to contrast our posterior draws of $\theta_{(q,s,k) i}$ for all $i$, either across time periods or at a fixed point in time, as these estimates account for league particulars such as strength of schedule. However, such a procedure would not scale to other sports or leagues where betting market data may be unavailable. Rather, we would prefer a metric that can be applied generally to any competitive scenario where paired comparison probabilities can be calculated.

To assess the equivalence of all teams in each league, we consider the likelihood that---given any pair of teams chosen at random---the better team wins, by simulating estimates of $p_{(q,s,k) ij}$ using posterior draws of team strength, home advantage, and game level error. For our purposes, we define the \emph{better} team to be the one, \emph{a priori}, with a higher probability of winning that game. If a contest has no inherent randomness (consider the Harlem Globetrotters), then the better team \emph{always} wins.\footnote{The Harlem Globetrotters are an exhibition basketball team that plays hundreds of games in a year, rarely losing.} Conversely, if game-level variability is large relative to the difference in team strength, then even the inferior team might win nearly half the time.  

Using our posterior draws, we approximate the distribution of game-level probabilities between two randomly chosen teams using the following steps. Posterior draws from Model IHA are used.

Given sport $q$ with season length $K_{q}$, number of seasons $S_{q}$, and number of teams $t_{q}$, 

\begin{enumerate}
\item Draw season $\tilde{s}$ from $\left\{1, \ldots, S_{q} \right \}$, and week $\tilde{k}$ from $\left\{1, \ldots , K_{q} \right \}$.

\item Draw teams $\tilde{i}$ and $\tilde{j}$ from $\left \{1, \ldots , t_{q} \right \}$ without replacement.

\item Sample one posterior draw of team strength for $\tilde{i}$ and $\tilde{j}$, $\tilde{\theta}_{(q,\tilde{s},\tilde{k}) \tilde{i}}$ and $\tilde{\theta}_{(q,\tilde{s},\tilde{k}) \tilde{j}}$, respectively, from the posterior distributions of $\tilde{i}$ and $\tilde{j}$'s team strength estimates during season $\tilde{s}$ at week $\tilde{k}$. For simplicity, assume $\tilde{\theta}_{(q,\tilde{s},\tilde{k}) \tilde{i}} > \tilde{\theta}_{(q,\tilde{s},\tilde{k}) \tilde{j}}$. 

\item Sample one posterior draw of the HA, $\tilde{\alpha}_{q_0}$, from the posterior distribution of $\alpha_{q_0}$, as well as one posterior draw of team $\tilde{i}$'s home advantage, $\tilde{\alpha}_{(q)\tilde{i}*}$. 

\item Sample one posterior draw of the game-level variance parameter, $\tilde{\sigma}^2_{q, game}$, and draw a game-level error, $\tilde{\epsilon}_{q, game}$, from $\tilde{\epsilon}_{q, game} \sim N(0, \tilde{\sigma}_{q, game})$

\item Impute the simulated log-odds of $\tilde{i}$ beating $\tilde{j}$, $\text{logit}(\tilde{p}_{(q,\tilde{s},\tilde{k}) \tilde{i}\tilde{j}}) = \tilde{\alpha}_{q_0} + \tilde{\alpha}_{(q)\tilde{i}*} + \tilde{\theta}_{(q,\tilde{s},\tilde{k}) \tilde{i}} - \tilde{\theta}_{(q,\tilde{s},\tilde{k}) \tilde{j}} + \tilde{\epsilon}_{q, game}$. 

\item Transform $\text{logit}(\tilde{p}_{(q,\tilde{s},\tilde{k}) \tilde{i}\tilde{j}})$ into a probability to obtain a simulated estimate, $\tilde{p}_{q, sim}$, where  $\tilde{p}_{q, sim} = \tilde{p}_{(q,\tilde{s},\tilde{k}) \tilde{i}\tilde{j}}$
\item Repeat the above steps $n_{sim}$ times to obtain $\mathbf{\tilde{p}_{q}} = \left\{\tilde{p}_{q, 1}, \ldots, \tilde{p}_{q, n_{sim}}\right\}$.
\end{enumerate}

For each $q$, we simulated with $n_{sim} = 1000$. Additionally, to remove the effect of each league's HA on simulated probabilities, we repeated the process fixing $\tilde{\alpha}_{q_0} = \tilde{\alpha}_{(q)\tilde{i}*} = 0$ for each league to reflect game probabilities played in absence of a home advantage.

\begin{knitrout}
\definecolor{shadecolor}{rgb}{0.969, 0.969, 0.969}\color{fgcolor}\begin{figure}
\includegraphics[width=\maxwidth]{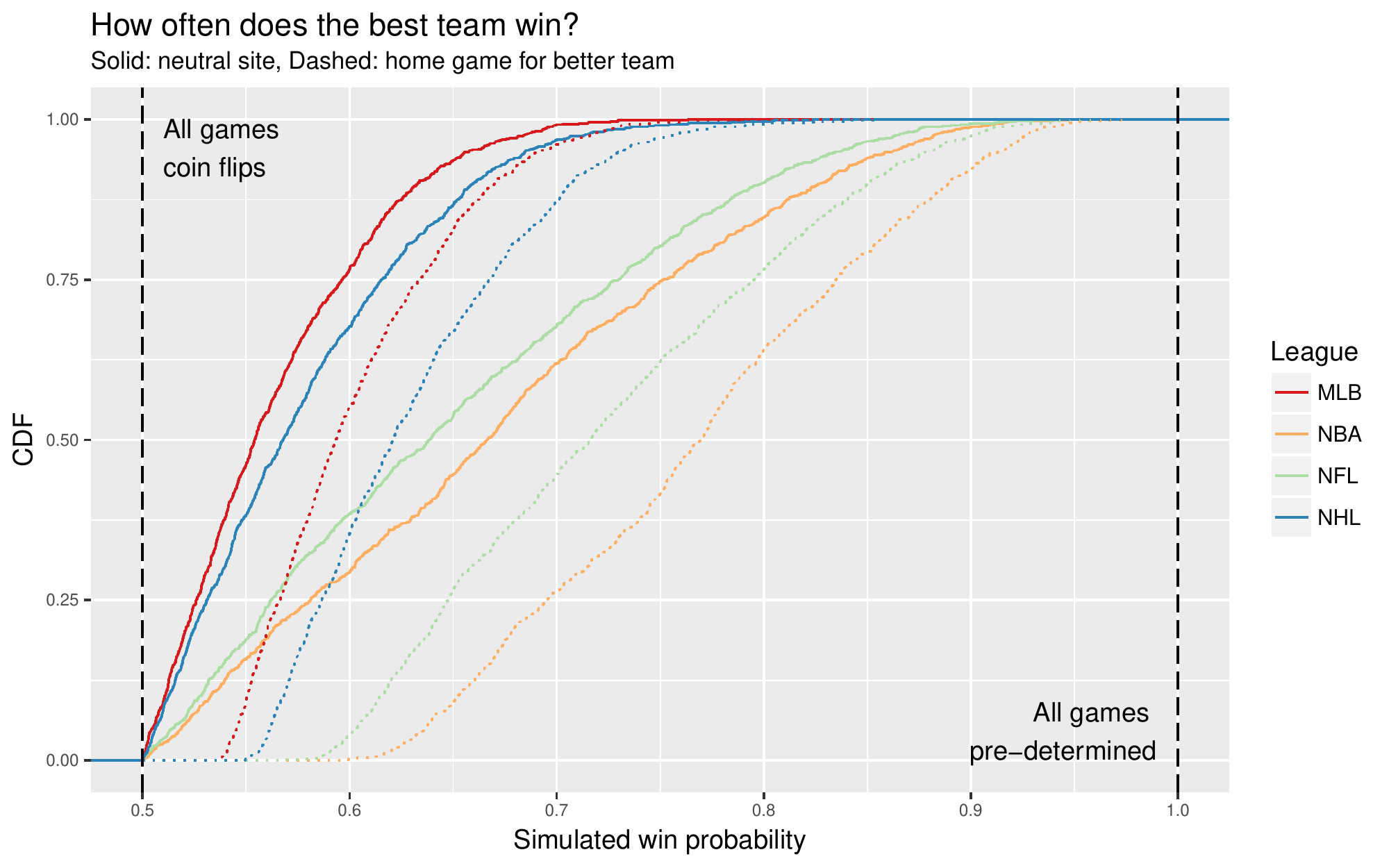} \caption[Cumulative distribution function (CDF) of 1000 simulated game-level probabilities in each league, for both neutral site and home games, with the better team (on average) used as the reference and given the home advantage]{Cumulative distribution function (CDF) of 1000 simulated game-level probabilities in each league, for both neutral site and home games, with the better team (on average) used as the reference and given the home advantage.}\label{fig:BestWin}
\end{figure}

\end{knitrout}

Figure~\ref{fig:BestWin} shows the cumulative distribution functions (CDFs) for each set of probabilities in each league. The median probability of the best team winning a neutral site game is highest in the NBA (67\%), followed in order by the NFL (64\%), NHL (57\%), and MLB (56\%). The spread of these probabilities are of great interest. Nearly every simulated MLB and NHL game played at a neutral site is less than a 3:1 proposition with respect to the best team winning (75\%). Meanwhile, roughly 27\% of NBA and 20\% of NFL neutral site match-ups are greater than this 3:1 threshold. 

Factoring in each league's home advantage works to exaggerate league-level differences. When the best team plays at home in the NBA, it is always favored to win at least 60\% of the time, with the middle 50\% of games ranging from a 68\% probability to an 84\% probability. Meanwhile, even with a home advantage, it is rare that the best MLB team is \emph{ever} given a 70\% probability of winning, with the middle 50\% of games ranging from 57\% to 63\%. 

Finally, we use the CDFs displayed in Figure~\ref{fig:BestWin} to quantify the cumulative difference between each league's game-level probabilities and a league of coin flips by estimating the approximate area under each curve. Let $RegParity_q$ be our regular season parity measure, such that

$$RegParity_q = 2 \int_{0.5}^1 P(\mathbf{\tilde{p}_{q}} \leq x) dx \, ,$$

\noindent where we multiply by 2 in order to scale so that $0 \leq RegParity_q \leq 1$, where 1 represents complete parity (every game a coin flip) and 0 represents no parity (every game outcome pre-determined). 

For games with no home advantage, $RegParity_{\text{MLB}} = 0.87$, followed by the NHL (0.84), NFL (0.70), and NBA (0.66). When the best team has a home advantage, parity is again the greatest in the MLB (0.79), followed by the NHL (0.73), NFL (0.55), and NBA (0.47). These results suggest that when the best team is playing at home, the NBA is closer to a world where every game outcome is predetermined than to one where every game outcome is a coin flip. Meanwhile, even when giving the best team a HA, MLB game outcomes remain lightly-weighted coin flips. 

\subsubsection{Parity in postseason tournaments}

Notions of parity in the regular season influence which teams make the playoffs, but each league conducts a single-elimination postseason tournament with a different structure. To what extent do those structures mitigate or reinforce the parity levels discussed in the previous section? We address these questions using our team strength estimates.

First, we collect the $z \in \{8, 16\}$ teams with the highest average team strength estimates over the last four weeks of each season, in each sport. We then seed (in descending order of team strength, irrespective of division or conference) and simulate 1000 postseason tournaments, in which each round consists of a best-of-7-game series, with the higher seed having the home field advantage in each round. The results shown in Figure~\ref{fig:playoff-sims} (in the Appendix) confirm that the relationship between seed and tournament finish is strongest in the NBA and the NFL, and considerably weaker in MLB and the NHL. These findings accord with our regular season parity measures. 

Next, we construct a postseason tournament parity metric that acts as a pseudo-$R^2$. Let ${\bf F}=(F_{1},F_{2},\ldots,F_{z})$ be a $z$-dimensional random vector with the $d^{th}$ element indicating the round of tournament finish of the $d^{th}$ seed. \footnote{We note that $\mathbf{F}$ depends on the vector of team strengths.} That is, for the $d^{th}$-seeded team, $F_d=1$ indicates that team finished as tournament champions, $F_d=2$ implies that team finished as runners-up, and so on. In a $z$-team tournament in which the higher seeded team always wins (i.e. the seeds determine the finish), the vector ${\bf F}$ is constant with $F_1=1$, $F_2=2$, $F_3=F_4=3$, $F_5=F_6=F_7=F_8=4$, etc., and in general, $\Ex[F_d] = F_d =  \lceil \log_2{d} + 1 \rceil$ for $d = 1, 2, \ldots, z$.  In the other extreme, where the seed is irrelevant (i.e., all values of $\theta$ are equal and there is no home advantage), $\Ex[F_d] = \sum_{d=1}^z \frac{1}{z} \cdot \lceil \log_2{d} + 1 \rceil = f_z$, where $f_z$ is a constant that depends on $z$.  

We define a pseudo-$R^2$ as:
$$
PostParity_z = 1 - \frac{(E[{\bf F}]-f_{z} 1_{z})'(E[{\bf F}]-f_{z} 1_{z})}{\sum_{d=1}^{z} (\lceil \log_2{d} + 1 \rceil - f_{z})^2}\,,
$$




\noindent where $d=1, \ldots, z$ iterates over the seeds, $f_z$ is the seed-weighted expected finish round (e.g., 4.0625 for a 16-team tournament), and $1_{z}$ is a vector of ones of length $z$. A $PostParity_{z}$ value of 0 indicates that the higher seed always wins, while a $PostParity_{z}$ value of 1 occurs when all seeds have the same expected finish. In a 16-team, 7-game series tournament, the NBA and NFL's $PostParity_{16}$ values (0.43 and 0.51) lag far behind those of MLB and the NHL (0.88 and 0.85, respectively).

While these simulations force all sports to use the same postseason tournament format, reality is quite different. Accordingly, we simulate tournaments while varying the number of teams who qualify (8 or 16) as well as the length of each postseason series (selected odd numbers between 1 and 75). Figure~\ref{fig:playoff-series} allows us to compare values of $PostParity_z$ for different tournament structures across all four sports. While $PostParity_8$ and $PostParity_{16}$ values may not be directly comparable, we note that the 1-game series played in the NFL results in parity similar to the current MLB and NHL formats. This leaves the NBA alone as the sport whose postseason tournament most likely coronates the strongest regular season teams. Conversely, the playoff structure in MLB, which includes a single-game wild card play-in\footnote{We did not include the wild card game in our simulations.} and a 5-game division series, serves to undermine advantages conferred based on seed. In order to approach the level of parity (or lack thereof) of the NBA playoffs, MLB would have to switch to a 16-team tournament in which each round was approximately 75-game series. Conversely, in order to the achieve the level of parity in the other three sports, the NBA would have to reduce the number of playoff teams to 8, and play a single-game tournament.

\begin{knitrout}
\definecolor{shadecolor}{rgb}{0.969, 0.969, 0.969}\color{fgcolor}\begin{figure}
\includegraphics[width=\maxwidth]{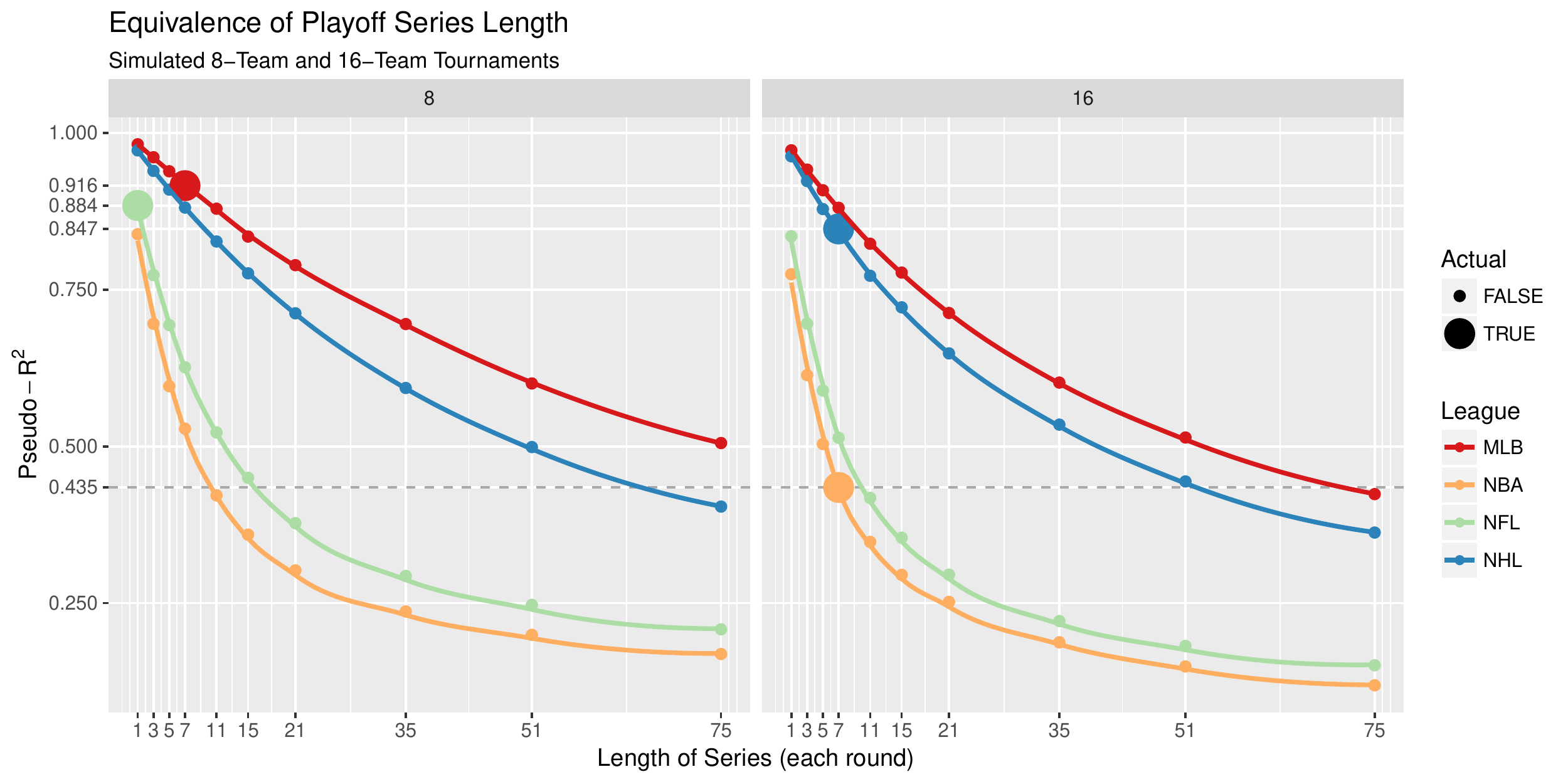} \caption[Parity measures for simulated playoff tournaments]{Parity measures for simulated playoff tournaments. Each line shows how our pseudo-$R^2$ parity metric changes as a function of tournament series length for both 8- and 16-team tournaments in each sport. We note that in order for MLB to achieve the same lack of parity as the NBA, it would have to play 75-game series in a 16-team tournament. Conversely, the NBA would have to switch to an 8-team, single-game tournament to match the parity of the other three sports. }\label{fig:playoff-series}
\end{figure}

\end{knitrout}

Postseason parity cuts both ways: a tournament in which the higher seeds always win is potentially less interesting, but a tournament in which seeds don't matter might compromise the competitiveness of late-season games for playoff teams. This represents a philosophical choice for commissioners. The NBA has clearly chosen a postseason structure that---relative to other sports---largely ensures that the best teams will win most of the time. We suspect that this arrangement is comforting for players and team executives, since the hard work of building a good team is remunerated with postseason success. On the other hand, early-round games may suffer from lack of interest, since fans may consider the outcomes predetermined. Conversely, MLB (and to a slightly lesser extent the NFL and NHL) postseason structure serves to maximize fan interest (recall the outcome uncertainty hypothesis), while offering few postseason rewards (other than entry) for regular season success. This may be an acceptable trade-off, since the regular season is so long and relatively few teams make the playoffs. Still, it may be profoundly frustrating to players and team executives.

\section{Conclusion}

\subsection{Summary}

We propose a modified Bayesian state-space framework that can be used to estimate both time-varying strength and variance parameters in order to better understand the underlying randomness in competitive organizations. We apply this model to the NBA, NFL, NHL, and MLB.

Our first finding relates to the relative equivalence of the four leagues. At a single point in time, team strength estimates diverge substantially more in the NBA and NFL than in the NHL and MLB. In the latter two leagues, contests between two randomly chosen teams are closer to a coin-flip, in which each team has a reasonable shot at winning. Understanding this underlying randomness would appear to be crucial for decision makers in these leagues. At critical moments in a team's evolution, such as the a trade deadline, free agency period, or the decision to fire a coach, we recommend that team officials look past wins and losses to better understand team strength in the context of their league. As one easy example, it is insufficient to evaluate a baseball or hockey team based on their performance in the postseason alone, given that so many of those contests are nearly 50-50 outcomes. 

Our next finding relates to the relative equivalence of the home advantage in each league, with the NBA well ahead of the pack, with teams averaging a 62.0\% chance of winning versus a like-caliber opponent. We also show that the home advantage varies most significantly between venues within each of the NBA and the NHL. In the NBA, for example, the league's best team home advantage is worth a few wins per year, in expectation, over the league's worst home advantage. Moreover, with the exception of the Colorado Rockies, it is not clear that any MLB or NFL team has a statistically significant home effect. 

Finally, we identify that team strengths derived from $sequential$ fits are nearly as accurate for predictive purposes as the observed game probabilities, as judged by both links to future team performance and game-level outcomes. We conclude by using team strength draws to propose two parity metrics, one for regular season comparisons and another for postseason contrasts.

\subsection{Discussion}

There are several options for applying or extending our model. Generally, the conditions needed to apply our framework are minimal; only paired events, outcome probabilities, and some unit of time are needed.

As alternative examples in sports, comparisons between divisions of teams in the same organization (as in English soccer) or between the top leagues of the same sport (as in European soccer) would follow a similar structure to the one provided. Alternatively, in any sporting league, modeling the impact of structural changes (such as free agency, expansion or scoring system updates) would be straightforward to test by adding covariates to our models. Note that team sports are not required for our model to apply: a similar framework could assess the caliber of tennis players, whose relative strengths fluctuate over time both within and across seasons. Competitive balance questions within amateur sport (for example, conferences in NCAA football, or even across all intercollegiate sports) would follow a similar design. 

There are also several ways our model could generalize to other competitive spheres of life. Assessing player and team strength in the increasing popular (and visible) world of online gaming could be a future application. Online trivia leagues (e.g. the Learned League) also pit players organized into divisions by ability in head-to-head competition---their relative strengths could be modeled in our framework. Given that political elections have only one outcome, traditional prediction models are difficult to judge and calibrate. However, since our framework does not require outcomes, and expansive betting market data that tracks candidates' probabilities over time exists, applying our models to political elections is another possible extension. Comparisons in the volitility of candidate support over time, either between states, countries, or election cycles, may be feasible. 

Additionally, researchers of the NBA, NFL, NHL, and MLB could explore several hypotheses using our provided team strength estimates. One option would be to test how each league's scheduling quirks impact won-loss standings. For example, what is the consequence of the unbalanced schedule used in the NFL, relative to a balanced design? Given each league's schedule, how likely is it for the best team to qualify for postseason play? Finally, one could use time-varying estimates of team strength to consider the existence of tanking, in which teams---in order to secure a better draft position---are better off losing games later in the season. While this has been demonstrated in basketball using betting market data \citep{soebbing2013gamblers}, it would also be worth looking at tanking in other leagues, or if team interest in tanking corresponds to the perceived talent available in the upcoming draft.

Opportunities to improve our model are also plentiful. Both Models IHA and CHA make use of the logit transform with game-level probabilities. Although our posterior predictive checks do not seem to indicate that this assumption is unjustified, alternative transformations may be considered. We implemented a version of Model IHA that used the arcsin-sqrt transformation---the variance-stabilizing transformation for binomial proportions---and found nearly identical results with both game-level probability predictions and team strength estimates. 

To estimate predictions from the $sequential$ fits, we repeatedly applied our MCMC algorithm in each week. In place, sequential Monte Carlo samplers \citep{gilks2001following, del2006sequential} would have been more efficient. In the sports of soccer and hockey, one improvement would consider three-way lines that include the probability of a tie game. Specifically, soccer betting markets use a vector of probabilities (win, loss, tie). To account for these complexities, \cite{firth_2017} proposed a generalized Bradley-Terry model to simultaneously model both wins and draws, one that could likewise start with imputed game probability vectors. Finally, a comparison of team strengths estimated by our model, as well as those fit by \cite{glickman1998state} and \cite{koopmeiners2012comparison}, could more acutely identify the impact of using betting market data relative to point differential and won-loss outcomes.

To maintain consistency with the NFL's calendar, we considered time on a weekly basis; more refined approaches may be appropriate in other sports. As an example, investigation into starting pitchers in baseball---who change daily---could lead to novel findings. Additionally, another model specification could consider the possibility that time-varying estimates of team strength follow something other than an autoregressive structure. One alternative assumption, for example, is a stochastic volatility process \citep{glickman2001dynamic}. In this respect, our model can be considered a starting point for those looking to dig deeper in any sport without losing an ability to make cross-league or cross-sport comparisons.

\bibliographystyle{imsart-nameyear}
\bibliography{refs}

\begin{figure}[h]
\includegraphics[angle=90]{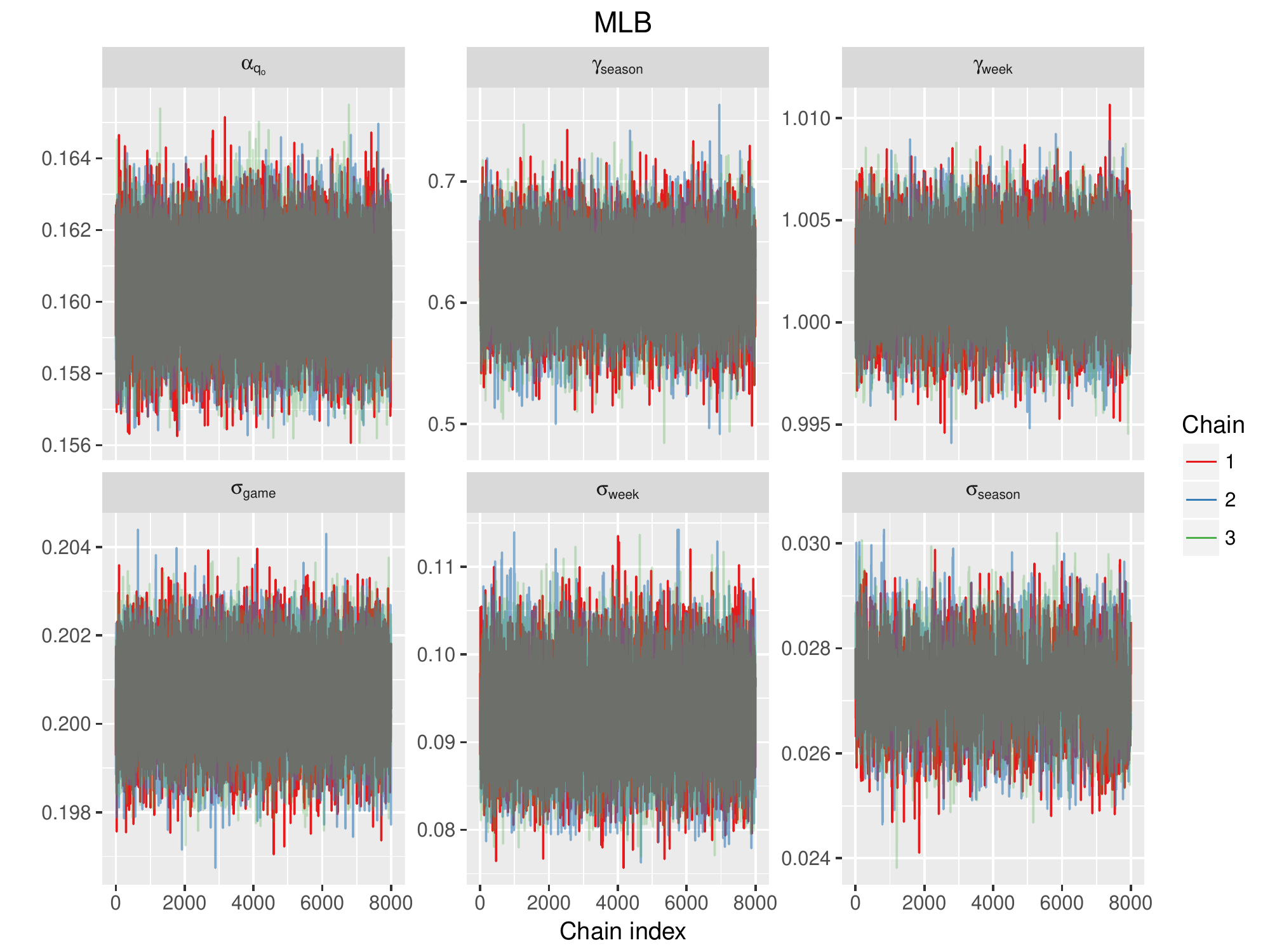}
\caption{Trace plots of MLB parameters\label{fig:MLBtrace}}
\end{figure}

\newpage

\begin{figure}[h]
\includegraphics[angle=90]{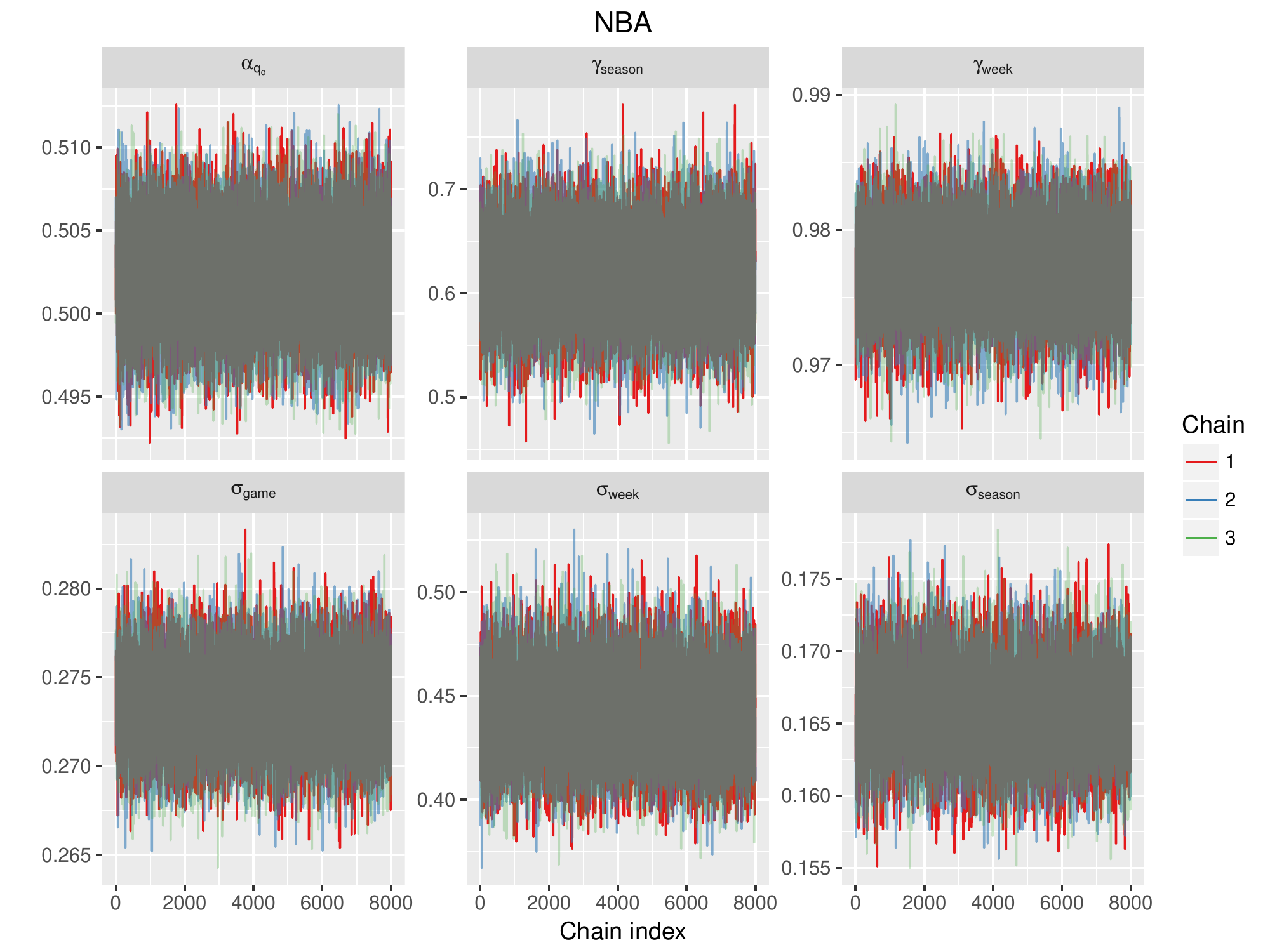}
\caption{Trace plots of NBA parameters\label{fig:NBAtrace}}
\end{figure}

\newpage

\begin{figure}[h]
\includegraphics[angle=90]{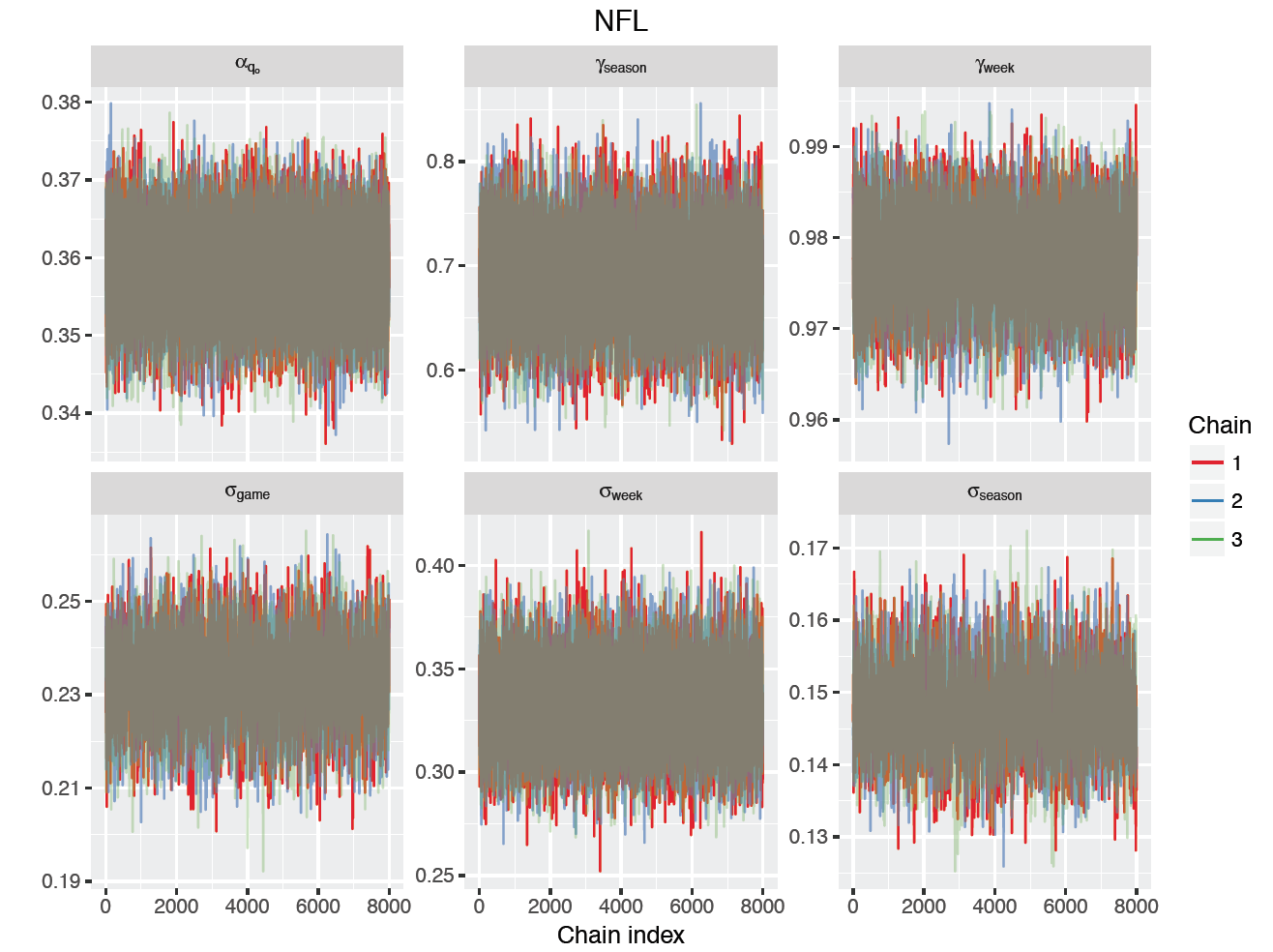}
\caption{Trace plots of NFL parameters\label{fig:NFLtrace}}
\end{figure}

\newpage 

\begin{figure}[h]
\includegraphics[angle=90]{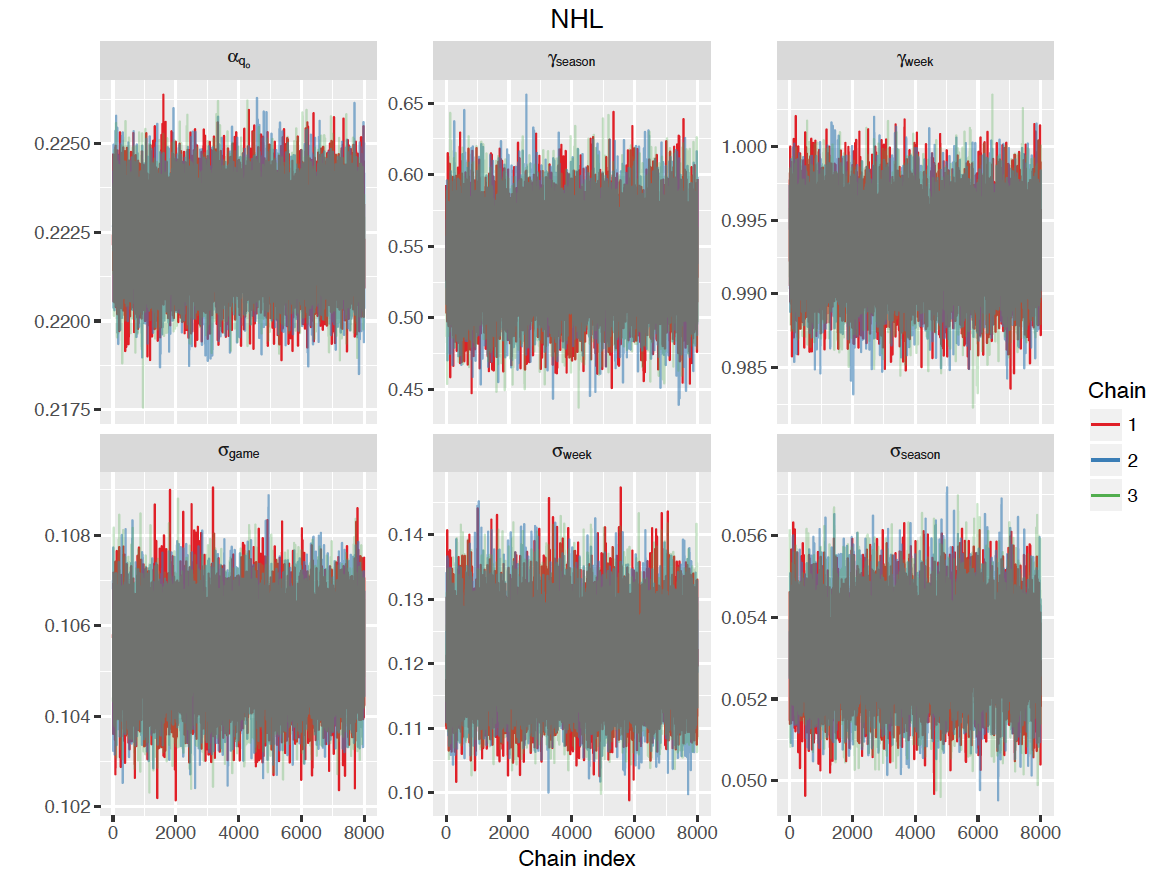}
\caption{Trace plots of NHL parameters\label{fig:NHLtrace}}
\end{figure}

\newpage

\begin{knitrout}
\definecolor{shadecolor}{rgb}{0.969, 0.969, 0.969}\color{fgcolor}\begin{figure}
\includegraphics[width=\maxwidth]{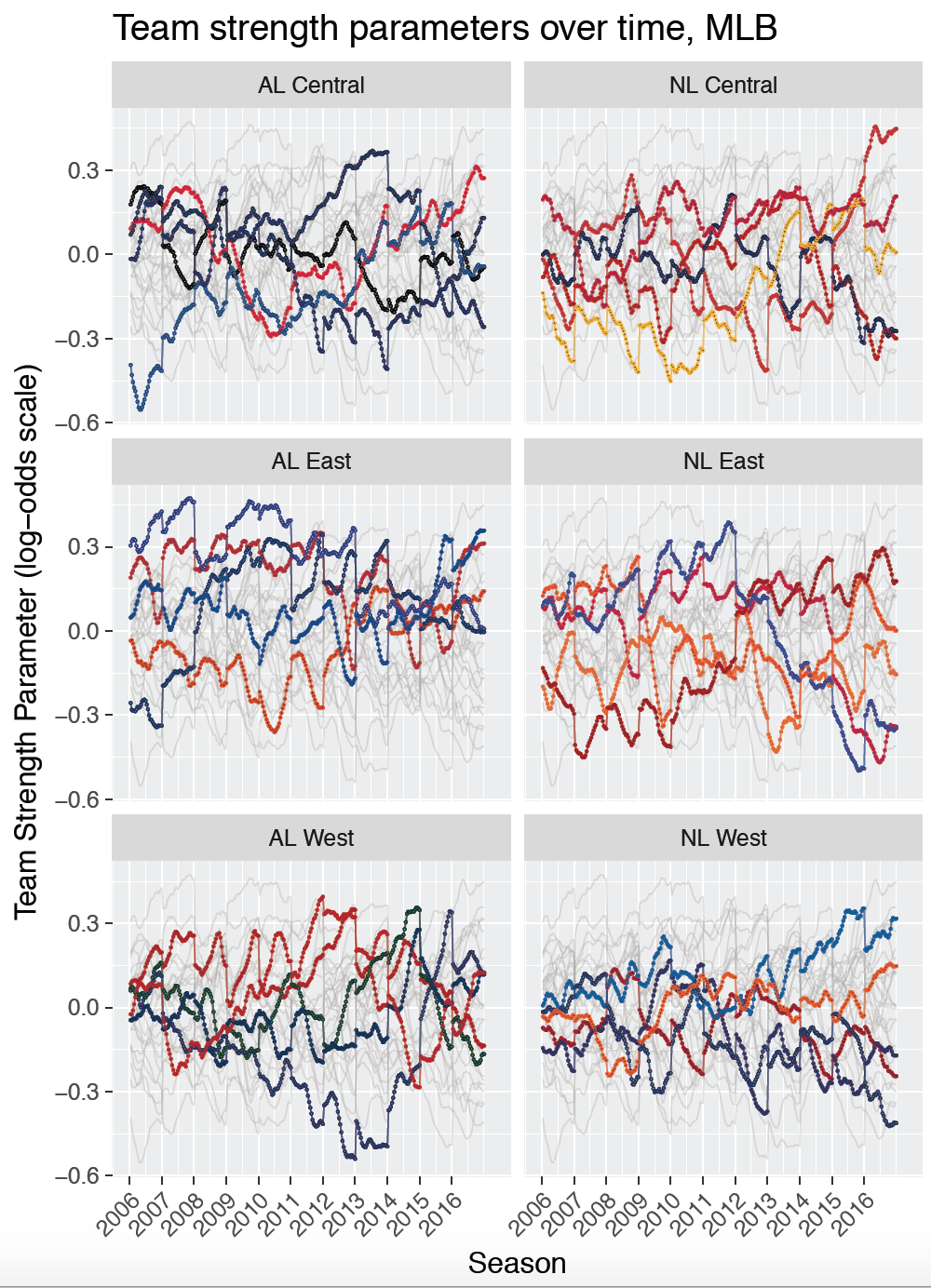} \caption[Team strength coefficients over time for Major League Baseball]{Team strength coefficients over time for Major League Baseball.}\label{fig:spaghetti-mlb}
\end{figure}

\end{knitrout}

\begin{knitrout}
\definecolor{shadecolor}{rgb}{0.969, 0.969, 0.969}\color{fgcolor}\begin{figure}
\includegraphics[width=\maxwidth]{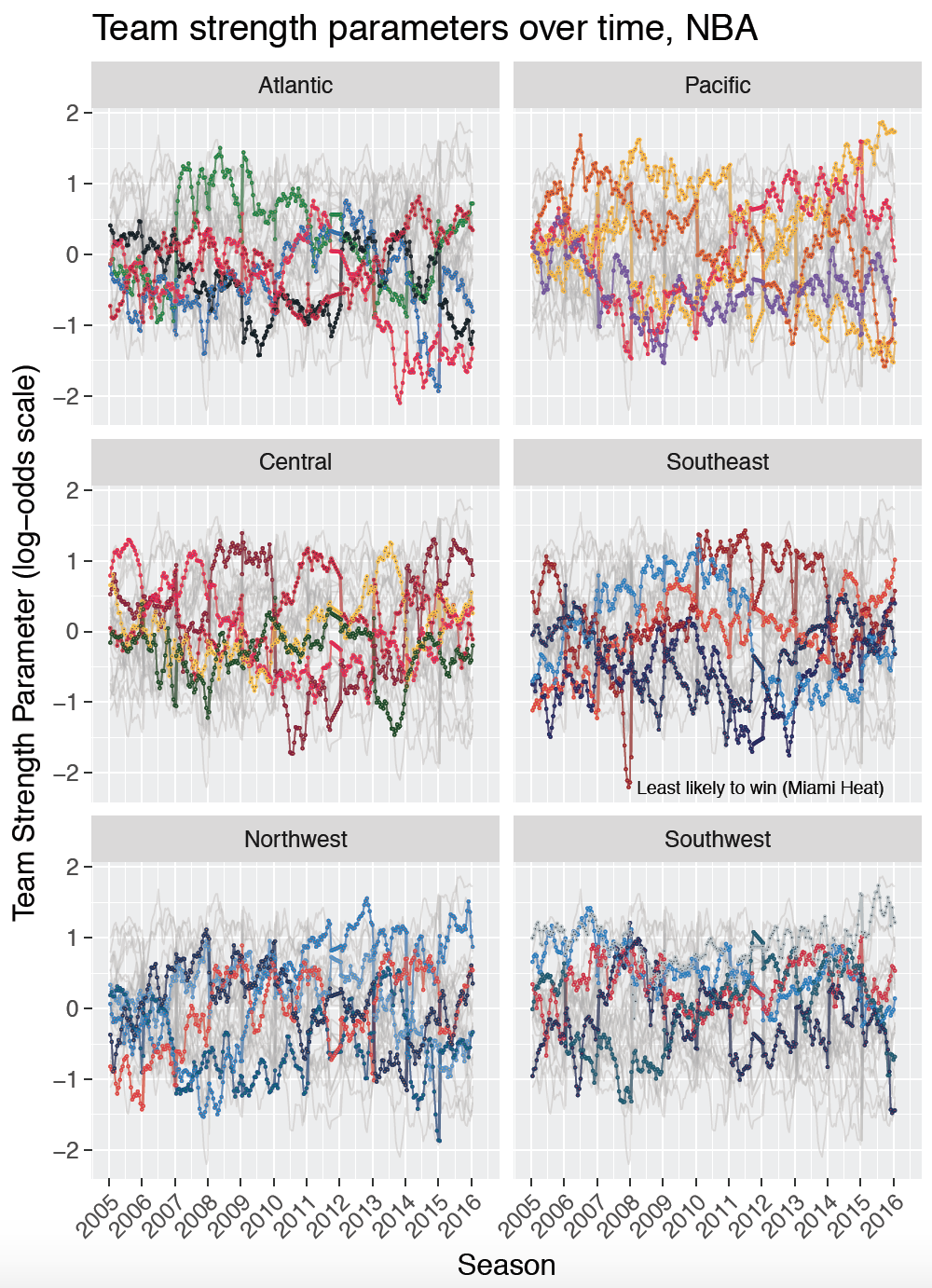} \caption[Team strength coefficients over time for the National Basketball Association]{Team strength coefficients over time for the National Basketball Association.}\label{fig:spaghetti-nba}
\end{figure}

\end{knitrout}

\begin{knitrout}
\definecolor{shadecolor}{rgb}{0.969, 0.969, 0.969}\color{fgcolor}\begin{figure}
\includegraphics[width=\maxwidth]{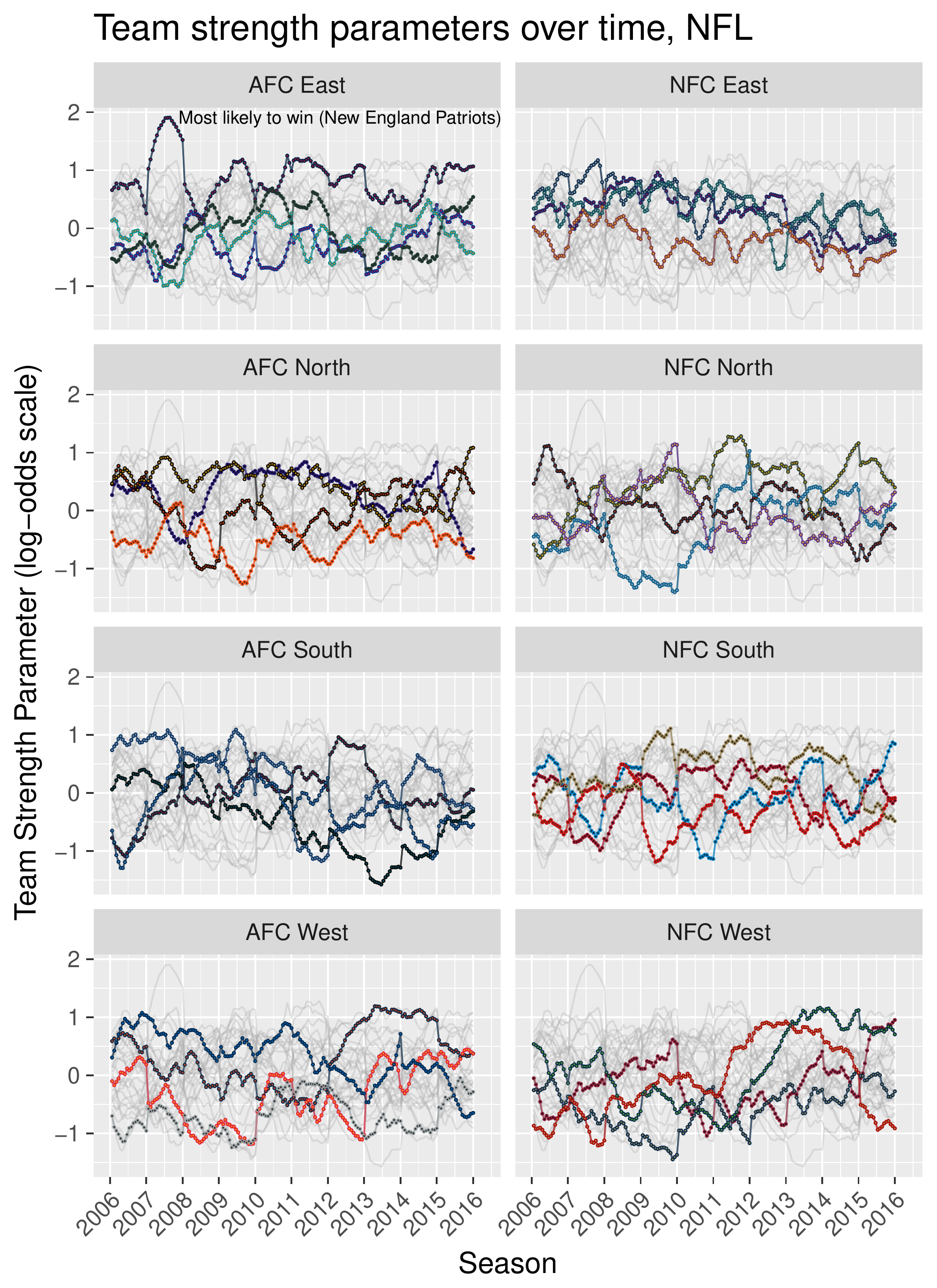} \caption[Team strength coefficients over time for the National Football League]{Team strength coefficients over time for the National Football League.}\label{fig:spaghetti-nfl}
\end{figure}

\end{knitrout}

\begin{knitrout}
\definecolor{shadecolor}{rgb}{0.969, 0.969, 0.969}\color{fgcolor}\begin{figure}
\includegraphics[width=\maxwidth]{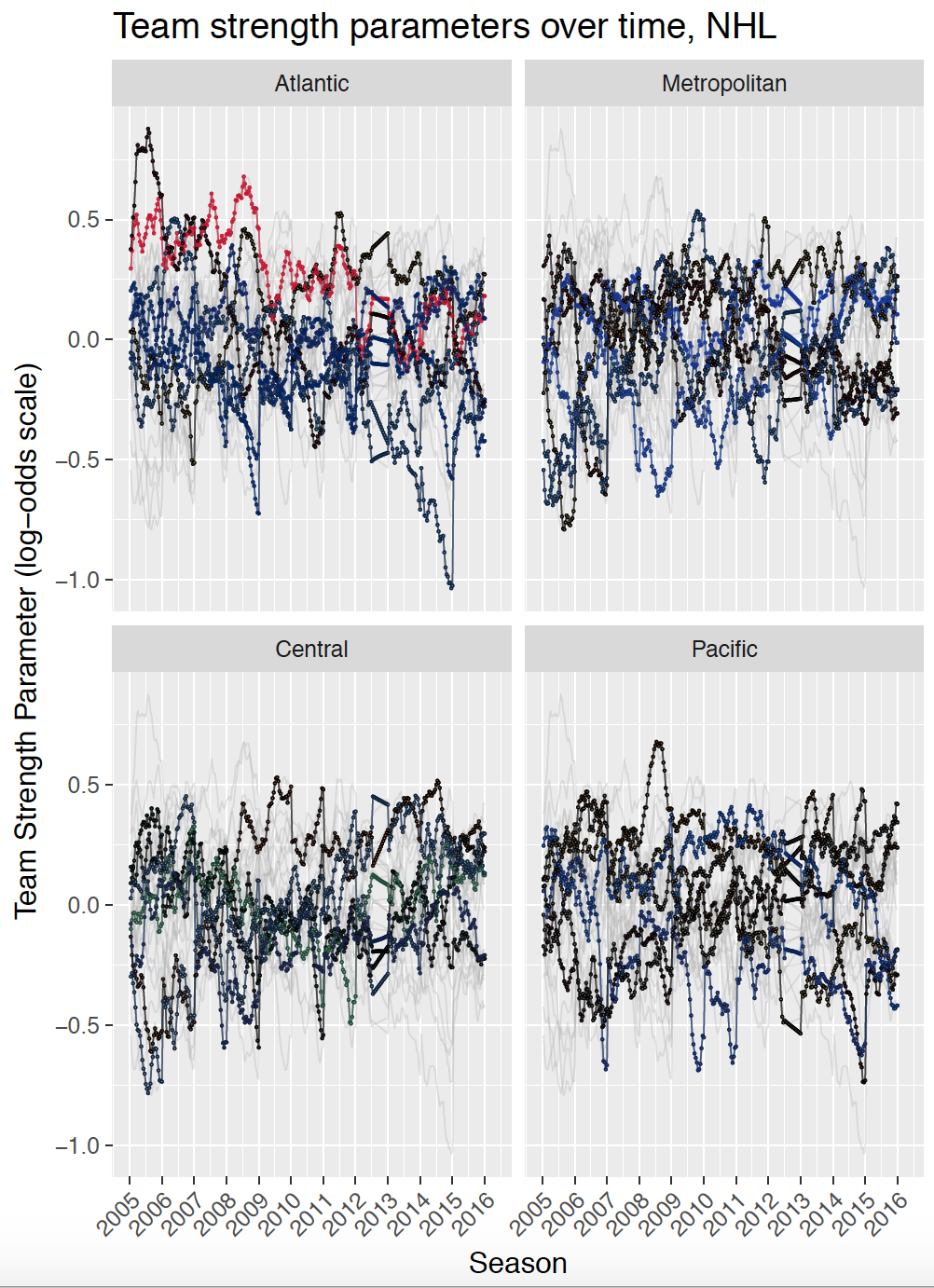} \caption[Team strength coefficients over time for the National Hockey League]{Team strength coefficients over time for the National Hockey League.}\label{fig:spaghetti-nhl}
\end{figure}

\end{knitrout}

\begin{knitrout}
\definecolor{shadecolor}{rgb}{0.969, 0.969, 0.969}\color{fgcolor}\begin{figure}
\includegraphics[width=\maxwidth]{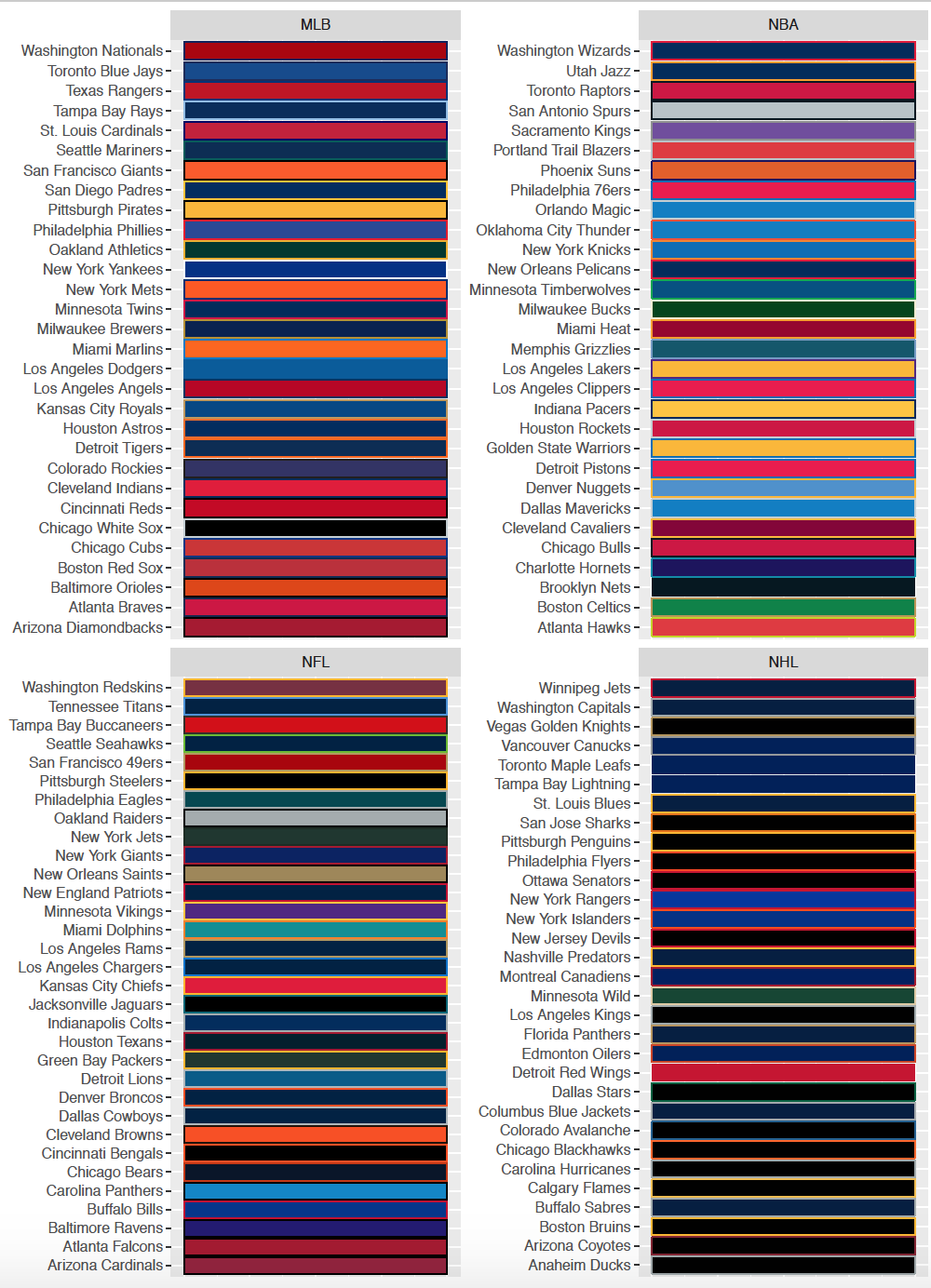} \caption[Team-color mappings used throughout the paper.~\citep{teamcolors}]{Team-color mappings used throughout the paper.~\citep{teamcolors}}\label{fig:teamcolors}
\end{figure}

\end{knitrout}

\begin{knitrout}
\definecolor{shadecolor}{rgb}{0.969, 0.969, 0.969}\color{fgcolor}\begin{figure}
\includegraphics[width=\maxwidth]{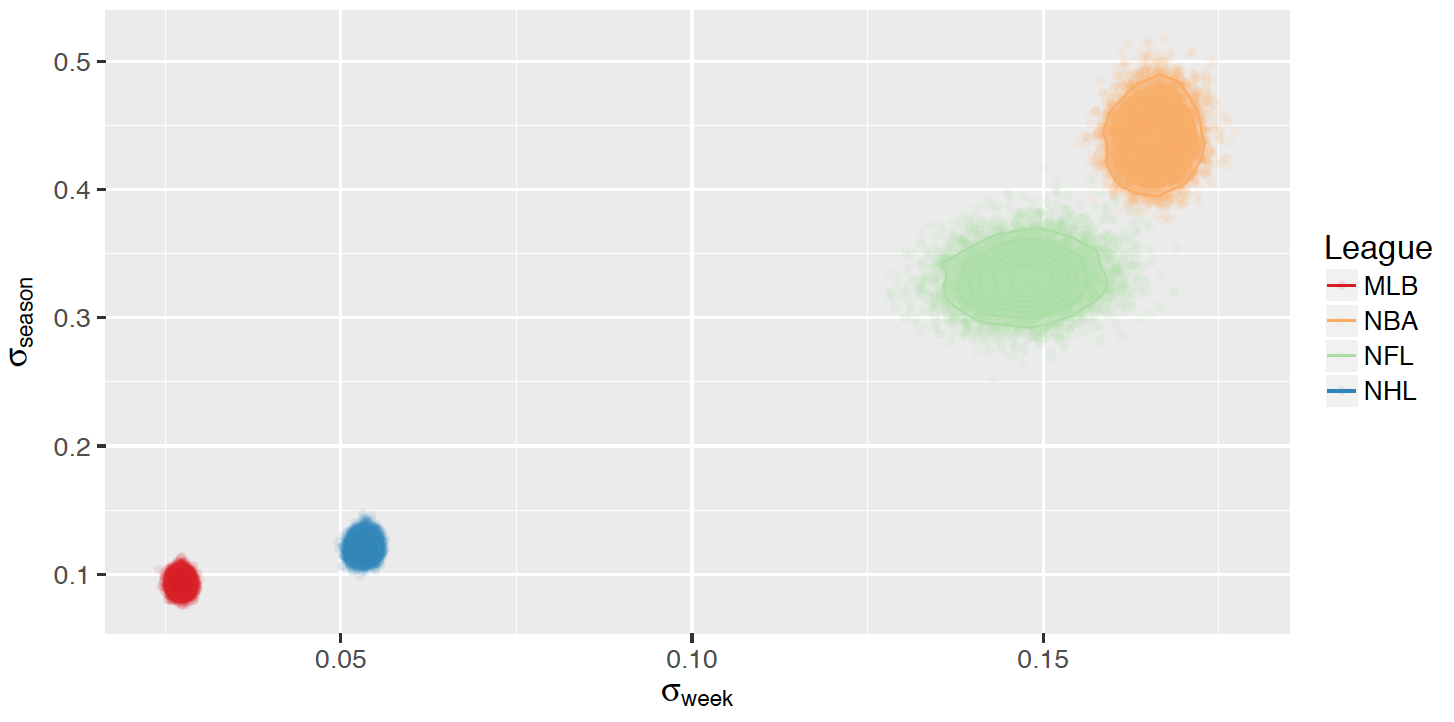} \caption[Contour plot of the estimated season-to-season and week-to-week variability across all four major sports leagues]{Contour plot of the estimated season-to-season and week-to-week variability across all four major sports leagues. By both measures, uncertainty is lowest in MLB and highest in the NBA.}\label{fig:contourSigma}
\end{figure}

\end{knitrout}

\begin{knitrout}
\definecolor{shadecolor}{rgb}{0.969, 0.969, 0.969}\color{fgcolor}\begin{figure}
\includegraphics[width=\maxwidth]{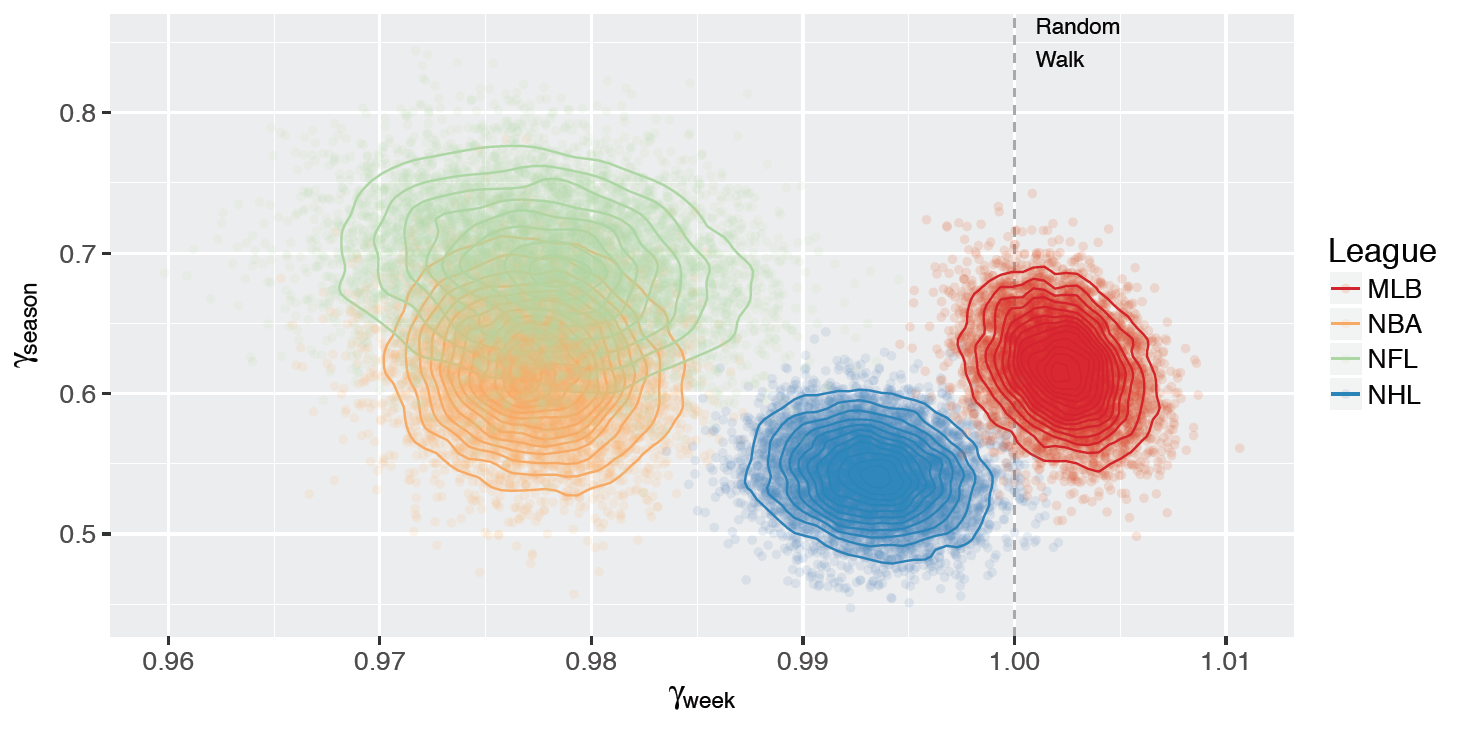} \caption[Contour plot of the estimated season-to-season and week-to-week autoregressive parameters across all four major sports leagues]{Contour plot of the estimated season-to-season and week-to-week autoregressive parameters across all four major sports leagues. }\label{fig:contourGamma}
\end{figure}

\end{knitrout}

\begin{knitrout}
\definecolor{shadecolor}{rgb}{0.969, 0.969, 0.969}\color{fgcolor}\begin{figure}
\includegraphics[width=7.5in,angle=90]{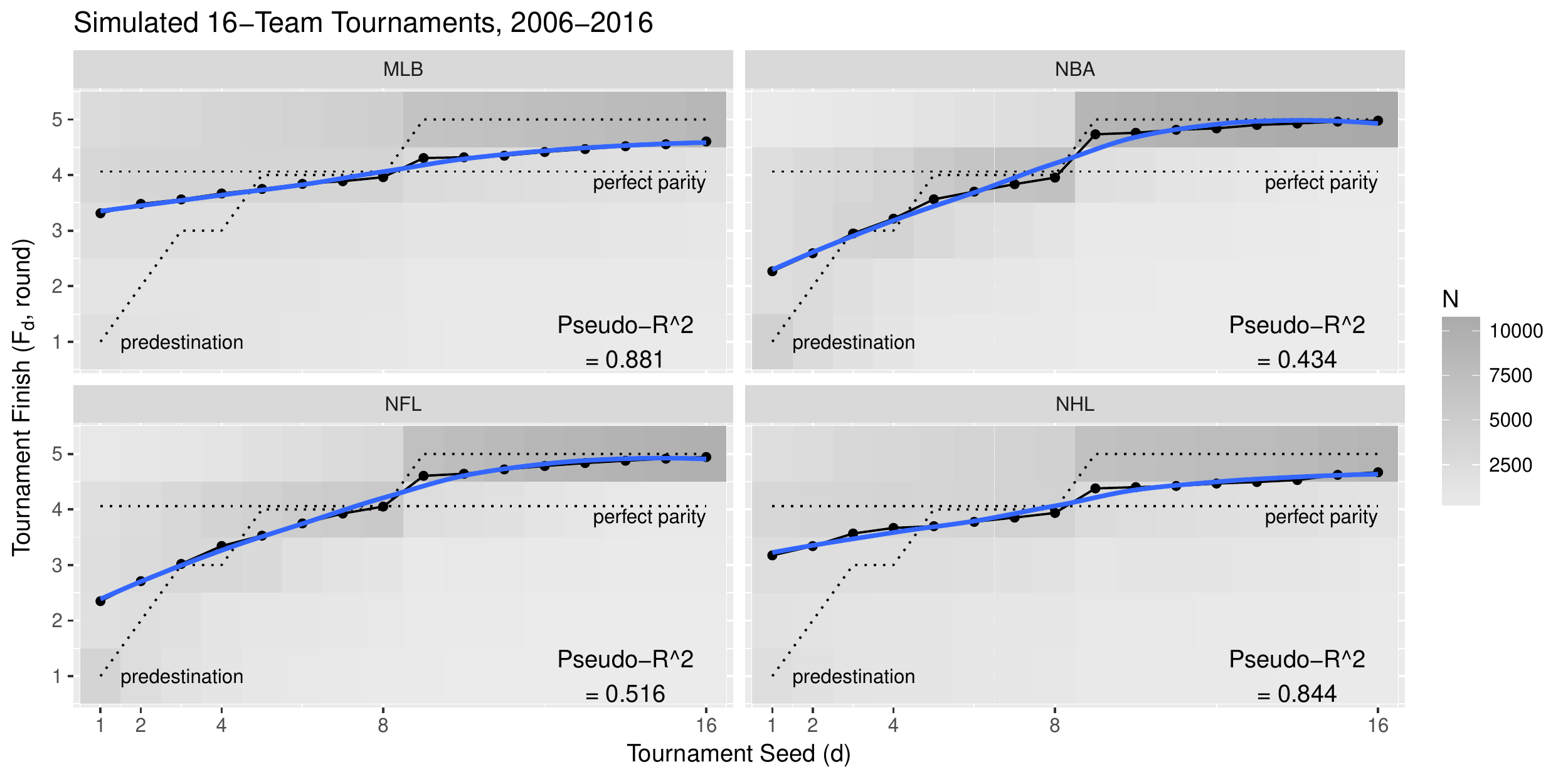} \caption[Relationship between seed and finish in simulated 16-team, 7-game series playoff tournaments]{Relationship between seed and finish in simulated 16-team, 7-game series playoff tournaments. One thousand tournaments were simulated for each sport in each year. The horizontal dotted gray line represent how the tournaments would play out with perfect parity, while the stepped gray line represents tournaments that play out in perfect accordance with seed. }\label{fig:playoff-sims}
\end{figure}

\end{knitrout}

\phantom{xxxx}

\vspace{3cm}

\begin{center}
{\Large  {\bf Supplementary Materials for \\ 

\vspace{2cm}

``A unified approach to understanding randomness in sport"}}
\end{center}

\newpage





\end{document}